\documentclass[12pt,preprint]{aastex}

\slugcomment{Accepted for Publictation in the Astronomical Journal}

\shorttitle{IR Spectroscopy of Comet SW3}
\shortauthors{Sitko et al.}

\begin{document}

\title{Infrared Spectroscopy of Comet 73P/Schwassmann-Wachmann 3 using the \textit{Spitzer Space Telescope}}

\author{Michael L. Sitko\altaffilmark{1,2,3}}
\affil{Space Science Institute, 475 Walnut St., Suite 205, Boulder, CO 80301}
\email{sitko@spacescience.org}

\author{Carey M. Lisse}
\affil{Applied Physics Lab, Johns Hopkins University, 11100 Johns Hopkins Road, Laurel, MD 20723}
\email{carey.lisse@jhuapl.edu}

\author{Michael S. Kelley}
\affil{Dept. of Astronomy, University of Maryland, College Park, MD 20742}
\email{msk@astro.umd.edu}

\author{Elisha F. Polomski}
\affil{Dept. of Physics and Astronomy, University of Wisconsin, Eau Claire, WI 54702}
\email{epolomsk@uwec.edu}

\author{David K. Lynch\altaffilmark{2}, Ray W. Russell\altaffilmark{2}}
\affil{The Aerospace Corporation, Los Angeles, CA 90009}
\email{David.K.Lynch@aero.org, Ray.W.Russell@aero.org}

\author{Robin L. Kimes}
\affil{Dept. of Physics, University of Cincinnati, Cincinnati OH 45221}
\email{rlk824@gmail.com}

\author{Barbara A. Whitney, Michael J. Wolff}
\affil{Space Science Institute, 475 Walnut St., Suite 205, Boulder, CO 80301}
\email{bwhitney@wisc.edu,mjwolff@spacescience.org}

\and

\author{David E. Harker}
\affil{Center for Astrophysics and Space Science, University of California at San Diego, La Jolla, CA 92093}
\email{dharker@ucsd.edu}

\altaffiltext{1}{Guest observer, Spitzer Space Telescope}
\altaffiltext{2}{Guest Observer, Infrared Telescope Facility}
\altaffiltext{3}{Also Department of Physics, University of Cincinnati}

\begin{abstract}

We have used the \textit{Spitzer Space Telescope} Infrared Spectrograph (IRS) to observe the 5-37~\micron{} thermal emission of comet 73P/Schwassmann-Wachmann 3 (SW3), components B and C. We obtained low spectral resolution (R$\sim$100) data over the entire wavelength interval, along with images at 16 and 22 ~\micron{}. These observations provided an unprecedented opportunity to study nearly pristine material from the surface and what was until recently the interior of an ecliptic comet - cometary surface having experienced only two prior perihelion passages, and including material that was totally fresh. The spectra were modeled using a variety of mineral types including both amorphous and crystalline components. We find that the degree of silicate crystallinity, $\sim$35\%, is somewhat lower than most other comets with strong emission features, while its abundance of amorphous carbon is higher. Both suggest that SW3 is among the most chemically primitive solar system objects yet studied in detail, and that it formed earlier or farther from the sun than the bulk of the comets studied so far. The similar dust compositions of the two fragments suggests that these are not mineralogically heterogeneous, but rather uniform throughout their volumes. The best-fit particle size distribution for SW3B has a form dn/da$\sim$a$^{-3.5}$, close to that expected for dust in collisional equilibrium, while that for SW3C has dn/da$\sim$a$^{-4.0}$, as seen mostly in active comets with strong directed jets such as C/1995 O1 Hale-Bopp. The total mass of dust in the comae plus nearby tail, extrapolated from to the field of view of the IRS peakup image arrays, is 3-5 x 10$^{8}$ kg for B and 7-9 x 10$^{8}$ kg for C. Atomic abundances derived from the spectral models indicates a depletion of O compared to solar photospheric values, despite the inclusion of water ice and gas in the models. Atomic C may be solar or slightly sub-solar, but its abundance is complicated by the potential contribution of spectrally featureless mineral species to the portion of the spectra most sensitive to the derication of the C abundance.  We find a relatively high bolometric albedo, $\sim$0.13 for the dust, considering the large amount of dark carbonaceous material, but consistent with the presence of abundant small particles and strong emission features.

\end{abstract}

\section{Introduction}

A major goal of planetary science is to develop a theoretical framework that is capable of reproducing the characteristics of our solar system today. Comets form a direct link to the physical and chemical processing that occurred when our solar system was was only a few Myr old (0.02\% of its present age). These primitive materials consist of ices of abundant volatile molecules (water, methane, ammonia, carbon monoxide, carbon dioxide), and dust - highly refractory inorganic materials (rock-forming minerals), as well as organic materials of generally intermediate volatility. Of particular interest are the mineralogy, size distribution, and degree of crystallinity of the dust. These provide key information about the condensation of mineral types from the young solar nebula, the  dust-gas chemistry, as well as possible large-scale transport within the nebula. 

\subsection{Origin of Comets and their Dust Material}

Comets fall into two broad classes, based on their orbital dynamics. The nearly isotropic comets (NICs, previously called dynamically-new comets and Oort cloud comets) are those entering the inner solar system from nearly isotropic directions on highly eccentric orbits. They are thought to have been formed and ejected from the Jupiter-Uranus region, and are currently stored in the Oort cloud. The ecliptic comets (ECs), by contrast, are those with low orbital eccentricities  and with inclinations close to the ecliptic, and which include objects significantly affected by the gravity of Jupiter, so that they are in resonant or near resonant orbits with the giant planet, the Jupiter-family comets (JFCs). They are commonly believed to have formed further away from the Sun than the NICs were, near the outer edge of the nebula/disk, and they might be expected to possess distinctly different compositions than the Oort cloud objects. Dynamical studies suggest that the ECs as well as many of the intermediate Halley-family  comets (periodic non-ECs) and possibly some  NICs come from the trans-Neptunian scattered disk, a population still exhibiting these eccentricity and inclination patterns \citep{dl97,fgb04,lddg06,vm08}. Simulations of the orbital evolution of the planets indicate that significant migration (and possibly even positional swapping by planets \citep{tsiganis05}) may have occurred during the time of comet formation, so that the distinction between the location of formation of the two comet populations may be blurred. 

The composition of the minerals that make up cometary  dust will be a mixture of pre-solar grains that are unchanged remnants of  interstellar medium (ISM), material that accreted into  the molecular cloud that the Sun formed in, materials that condensed within the cloud and proto-solar  nebula itself, and material which has undergone a sequence of condensation and evaporation and annealing  in the circulating proto-planetary disk formed after the nebula collapsed to form the young Sun and its accretion disk. The material that condenses (or re-condenses) in the nebula will be subject to a complex set of gas-grain (and gas-gas) interactions which can produce a radial stratification in the condensation sequence of minerals \citep{gail98}. Because of this, it is possible that comets forming at different radial distances may have inherited different mineralogies. 

It is known that the grains in the interstellar medium have a very low degree of crystallinity, only $\sim$1\% \citep{ciska04,ciska05}. This contrasts to the much higher values observed in comets - exceeding 50\% in some cases \citep{har02,har04,lisse07}. Where, when, and how this transformation occurs is still debated, but annealing and sublimation/condensation in the hottest part of the solar nebula is implicated. Spatially resolved spectroscopy of the silicate band of dust in young stellar disk systems indicates that the inner few AU have larger and more crystalline silicates than the more distant regions \citep{vb03}, suggesting production in the inner regions and radial transport outward.

A related aspect of the formation and processing of dust in the solar nebula is the availability of elemental species. Even the most ``primitive'' meteorites, the CI carbonaceous chondrites, are depleted in carbon and oxygen compared to solar photospheric values \citep{lodders09a,lodders09b}.  A similar pattern is seen in the mass spectrometer data on 1P/Halley obtained with the \textit{Giotto} and \textit{VEGA} spacecraft \citep{langevin87,jck88,jk91,lawler89}, which presumably sampled material condensed further from the Sun than the meteorites. These are critical elements for the formation of silicates, ices, and organics materials in the solar system. A radial gradient in the depletion pattern might provide clues about the condensation processes in the early solar nebula, and signal a lack of significant radial mixing. 

If there is a radial gradient in the characteristics of grain mineralogy and degree of crystallinity of the material in the early solar nebula, it would be manifested in differences between the grain characteristics of these two comet populations. Furthermore, if the grain characteristics of the comet-forming regions evolves while the cometesimals are growing and assembling, these bodies may be internally heterogeneous and possibly stratified. These differences should manifest themselves in differences among the grain characteristics of the comet populations. However, after many passages through the inner solar system, the surfaces of EC comets will become de-volatilized, processed, and may no longer represent the material they formed from. Furthermore, if the mineralogy and degree of crystallinity in the comet-building region was evolving as the comet nuclei were being assembled, then detection of inhomogeneities within the nucleus would provide a window on the temporal changes in the comet-building zone. Any stratification in grain properties would be difficult to detect from a mere snapshot study of its current (and processed) surface.

For these reasons it is important to sample the \textit{inside} of comet nuclei, particularly those of the ECs. One method is to artificially excavate beneath the surface layers, as was done by the \textit{Deep Impact (DI)} mission to comet 9P/Tempel 1 (hereafter T1). Another approach is to study comets that have recently fragmented, exposing their more pristine interiors for the first time. 

During the past decade or so, a number of ECs have been observed to fragment: D/1993 Shoemaker-Levy 9, 16P/Brooks 2, 51P/Harrington, 57P/du Toit-Neujmin-Delporte, 73P/Schwassmann-Wachmann 3 (hereafter SW3), 101P/Chernykh, 102P/Shoemaker-Holt 1, 174P/Echeclus, 205P/Giacobini, and P/2004 V5 (LINEAR-Hill) \citep{fernandez09}. Because disrupted comets may be responsible for the production of most of the dust in the zodiacal cloud \citep{nesvorny10}, understanding the physical and chemical properties of these comets  is important. As of 2006 there were no mid-IR spectra of a fragmented EC suitable for the necessary analysis. However SW3 survived a fragmentation event in 1995, and two orbits later was poised for a close approach to the Earth. This provided an unprecedented opportunity to study one of these objects in detail.

In this paper, we describe observations of the B and C fragments of SW3 using the Infrared Spectrograph (IRS) of the \textit{Spitzer Space Telescope}. Although \textit{Spitzer} was more distant from the comet than the Earth at its closest approach, its superior sensitivity, wavelength coverage, and lack of telluric spectral contamination provided us with the best opportunity to examine this important object in the mid-IR. Additional supporting ground-based visual, near-IR, and mid-IR observations that were useful for interpreting the \textit{Spitzer} data are described in the Appendix. We will also compare the results of the analysis of the \textit{Spitzer} observations with other independent mid-IR observations of SW3 \citep{har10} as well as other comets studied spectroscopically and using \textit{in situ} sampling techniques. 

In this study we will be able to:

$\bullet$ obtain information on the mineral content and degree of crystallinity on (potentially) the most pristine cometary material yet measured through remote sensing

$\bullet$ compare the abundances of the various minerals between two cometary fragments that represent different fractions of material formerly buried below the surface of the nucleus

$\bullet$ compare the atomic abundances of SW3 with other comets studies both spectroscopically and \textit{in sutu}, primitive meteorites, and the Sun, to investigate the pattern of elemental depletions among these solar system objects

\section{Observations \& Spectral Processing}

IRS observations of SW3-C and SW3-B were obtained on 17 March 2006 UT, and 17 April 2006 UT, respectively. For both objects, image maps were obtained using the Blue Peak-Up array ($\lambda=13.3-18.7 \mu$m, hereafter referred to as the 16~\micron{} array) and Red Peak-Up array ($\lambda=18.5-26.0 \mu$m, hereafter the 22~\micron{} array). The primary spectroscopic observations were obtained in the  ``stare'' mode with the low spectral resolution gratings. The various slits, spectral orders, wavelength coverage, and spectral resolutions $R = \lambda$/$\Delta\lambda$ of the IRS are listed in Table 1. In each case, the object was acquired with the 22~\micron{} array, then offset to the IRS slit for the spectral observations. 

SW3-C was observed at a heliocentric distance $r_{h}$=1.47 AU, and distance $\Delta$= 0.78 AU from \textit{Spitzer}. The 16 \& 22 ~\micron{} image maps indicated that this component consisted of a single fragment, with no evidence of smaller fragments within the field of view.

SW3-B was observed at $r_{h}$=1.19 AU, and a distance $\Delta$= 0.44 AU from \textit{Spitzer}. The observations were obtained within a day of the first set of \textit{Hubble Space Telescope} images \citep{hal06}, when the tail of SW3-B consisted of many sub-fragments\footnote{The 16~\micron{} \textit{Spitzer} IRS images have a point spread function (PSF) of 3.6 arcsec, which at the distance of SW3-B (0.44 AU) was too broad to detect the individual ``crumbs'' seen in the \textit{Hubble} images, which had superior spatial resolution due to its larger collecting mirror, shorter wavelength used (0.6~\micron{}), and closer distance at the time of observation (0.22 AU).}. \citet{ish09} reported detecting over 150 fragments using the  Subaru Suprime camera, in addition to the main intact body. The fragmentation that produced these pieces probably occurred around 1 April \citep{green06}, about 16 days before the \textit{Spitzer} observations.

Observations of SW3-G were also attempted, but failed to produce any data of sufficient quality for extraction and analysis. \textit{Hubble} images obtained less than 30 hours after the failed \textit{Spitzer} acquisition revealed G to have consisted entirely of a swarm of sub-fragments that were apparently the result of a major disintegration event a few days prior to the \textit{Hubble} observations \citep{hal06}. The failure was thus likely due to G being too diffuse, too faint, or both.

In addition to the low-resolution ``stare'' observations, observations were also made in the low-resolution ``spectral mapping'' mode to insure detection (albeit at a reduced signal-to-noise ratio) if the ephemeris was in error \footnote{Observations needed to be planned weeks before the initial recovery of the various fragments of the comet.}, to obtain spatial-spectral  information on the main components, and possibly measure the spectra of newly-shed fragments. A subset of the spectral maps were used to determine the slit loss correction factors for extracting the spectra from the ``stare'' observations (see Sec. 2.2). High spectral resolution measurements were also made in order to recover spectral information should the low-resolution spectra be saturated. In this paper we will concentrate solely on the low-resolution ``stare'' observations\footnote{The high resolution spectra do not include data shortward of 9.9 ~\micron{}, are noisier than the low resolution data, and (as of the time of the analysis) suffered severe order-to-order mis-matches for extended sources that made their merger problematic. However, the  extended source calibrations for the high-resolution data are improving and these data might be tractable in the near future.}

\subsection{Image Maps}

As part of the standard target acquisition process for \textit{Spitzer's} IRS, images are obtained from the peak-up arrays, are then used to center the target on the spectrograph slit. For SW3, these were saturated in the comae and not used for extracting science information. For this work, we instead used image maps constructed from a number of additional, short 6 sec exposure, dithered images we requested in each of the arrays. 

\begin{figure}
\plottwo{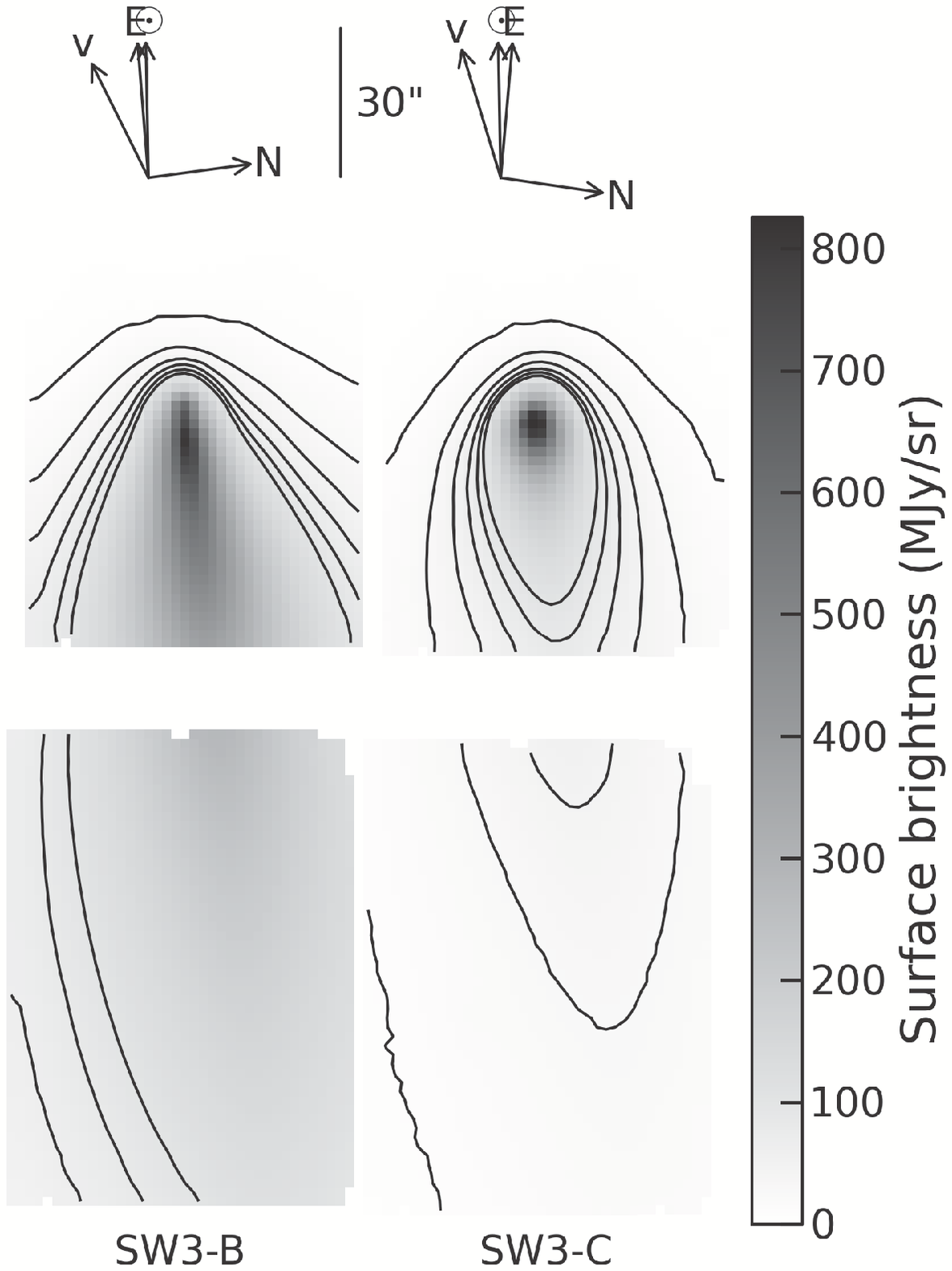}{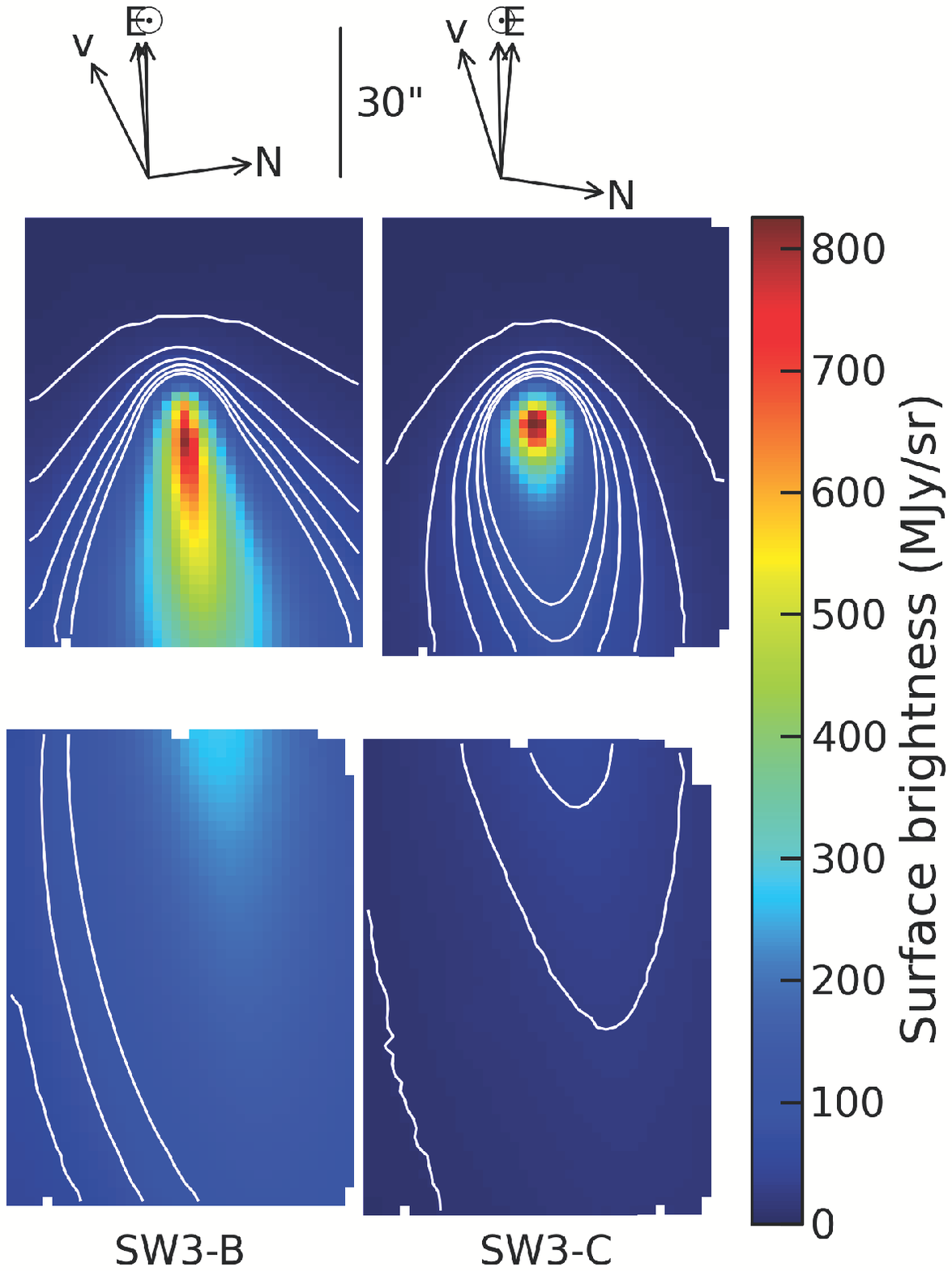}
\caption{Spitzer/IRS peak-up 16~\micron{} image maps of SW3-B
(left) and SW3-C (right).  Contours are plotted from 10, to 100 MJy sr${^-1}$
(inclusive), in linear intervals of 18 MJy sr$^{-1}$.  North, east, the comet's
projected heliocentric velocity vector (v), and the projected
direction of the sun ($\sun$) are shown above each fragment.  The
image scale is indicated at the top, and is the same for both image
maps.  The morphology of the SW3-C coma appears similar to that of the majority of comets, while the SW3-B coma is highly elongated and diffuse, most likely due to the recent fragmentation event preceding the observations.
See the electronic edition of the Journal for a color version 
of this figure.
\label{fig:puis}}
\end{figure}

The IRS peak-up array images were bad pixel masked, re-projected onto the
tangent plane, and mosaicked in the rest frame of the comet using the
MOPEX software \citep{mk05} version 18.2.0.  The 22~\micron{}
image maps covered clean backgrounds, but the 16~\micron{} images map did
not.  We estimate the 16~\micron{} background by measuring the
faintest edge of the images and multiplying by a factor of 0.97.
These edges are in the direction of the Sun, i.e., they do not
coincide with the comet tails.  The factor of 0.97 was derived by
comparing the faint, near-nucleus image edge of the 22~\micron{} images
to their background observations.  This method is likely accurate to
0.1~MJy sr$^{-1}$ pixel$^{-1}$.  The 16~\micron{} mosaics of fragments
B and C are presented in Fig.~\ref{fig:puis}.

\subsection{Extended Source Calibration and Spectral Extraction}

IRS spectra are calibrated with spectroscopic observations of point
sources.  The IRS slits are narrow with respect to the PSF of the
telescope, i.e., they do not encompass 100\% of the PSF at any
wavelength and the fraction of the PSF encompassed varies with
wavelength.  Thus, the low-resolution IRS slits block up to 35\% of
the flux of a point source at the longest wavelengths, but as little
as 17\% at the shortest wavelengths.  Because the IRS instrument is
calibrated with observations of unresolved stars, the slit-losses may
be ignored when working with calibrated observations of point sources.
Ignoring the slit-loss does, however, affect the spectral shape of
calibrated observations of sources larger than the PSF of the
telescope.  In the case of an infinitely extended source the
slit-losses are zero, but the calibration assumes losses equivalent to
a point source.  Therefore the resulting ``calibrated'' spectrum will
not accurately represent the spectral shape of the infinitely extended
source.

The thermal emission from comets is comprised of two parts: 1)
emission from the nucleus, generally spatially unresolved, and 2)
emission from the dust, which is a region of radially and azimuthally
varying surface brightness.  To properly flux-calibrate our comet
spectra, we compute the slit-losses for the observed surface
brightness distribution, and the orientations and sizes of each IRS
slit. No correction was applied for the presence of the nuclei of either fragment, as the emission from the nuclear surfaces was negligible compared to the emission from the huge surface areas of fine coma dust. (see Sec. 3.1). 

We began by measuring the spatial profile perpendicular to the
observed tail of each comet in the 5--7~\micron{} region of the IRS
data cubes built from the spectral maps using the CUBISM software \citep{smith07}. The 6~\micron{} region of the data cubes was chosen for
its high spatial resolution and good quality signal.  We fit photometric cuts
perpendicular to the tail with both Moffat and Gaussian functions, but
the Moffat functions  fit the images better.  From the best-fit Moffat
parameters we generated a model 6~\micron{} image of the comet at an
image scale of 0.1835~arcsec~pixel$^{-1}$, i.e., one-tenth the image
scale of the IRS peak-up arrays.  The super-resolution model image is
convolved with the synthetic IRS 16~\micron{} peak-up array PSF from a 300~K blackbody spectrum, to roughly approximate the the spectral shape of the thermal emission.  This PSF was
generated at an image scale of 0.1835 arcsec pixel$^{-1}$ with version
2.0 of the Tiny Tim/\textit{Spitzer} software\footnote{Tiny Tim/\textit{Spitzer} is
  available from the \textit{Spitzer} Science Center: \\
  http://ssc.spitzer.caltech.edu/}.  With the addition of a point
source component for fragment C, the surface brightness distribution
of the convolved super-resolution model images approximately match the
peak-up imaging observations.  The parameters of the Moffat fits, plus
the additional point source for fragment C, form the basis of our
model comets with which we derive the slit-losses of each comet.

\begin{figure}
\plotone{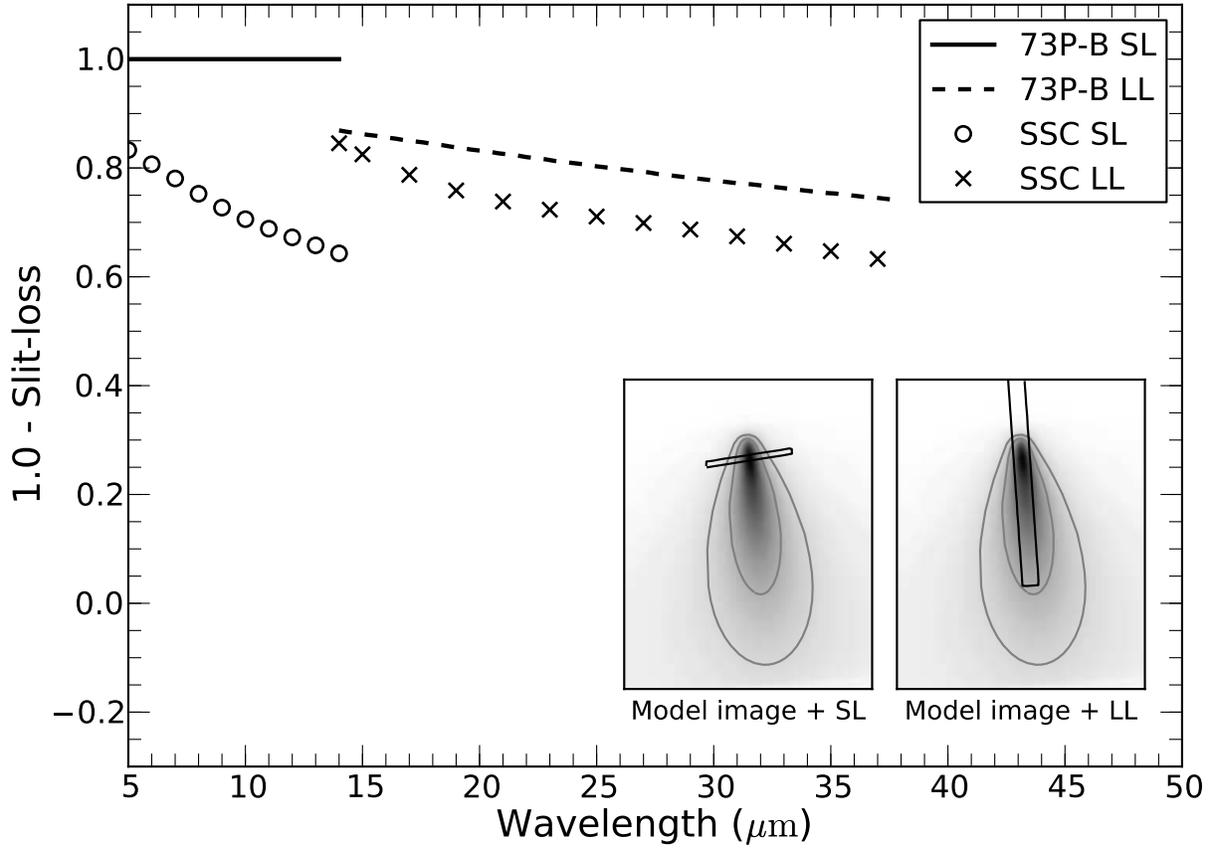}
\caption{Our computed slit-losses for our model fragment B surface
brightness distribution, and those of a point source as provided by
the Spitzer Science Center in the SPICE software's calibration files.
\textit{Inset:} Our model fragment B image convolved with a 16~\micron{} peak-up
array PSF.  Contours mark the locations of the IRS 5--14~\micron{}
(SL) and 14--38~\micron{} (LL) slits.  
\label{fig:slcfB}}
\end{figure}

\begin{figure}
\plotone{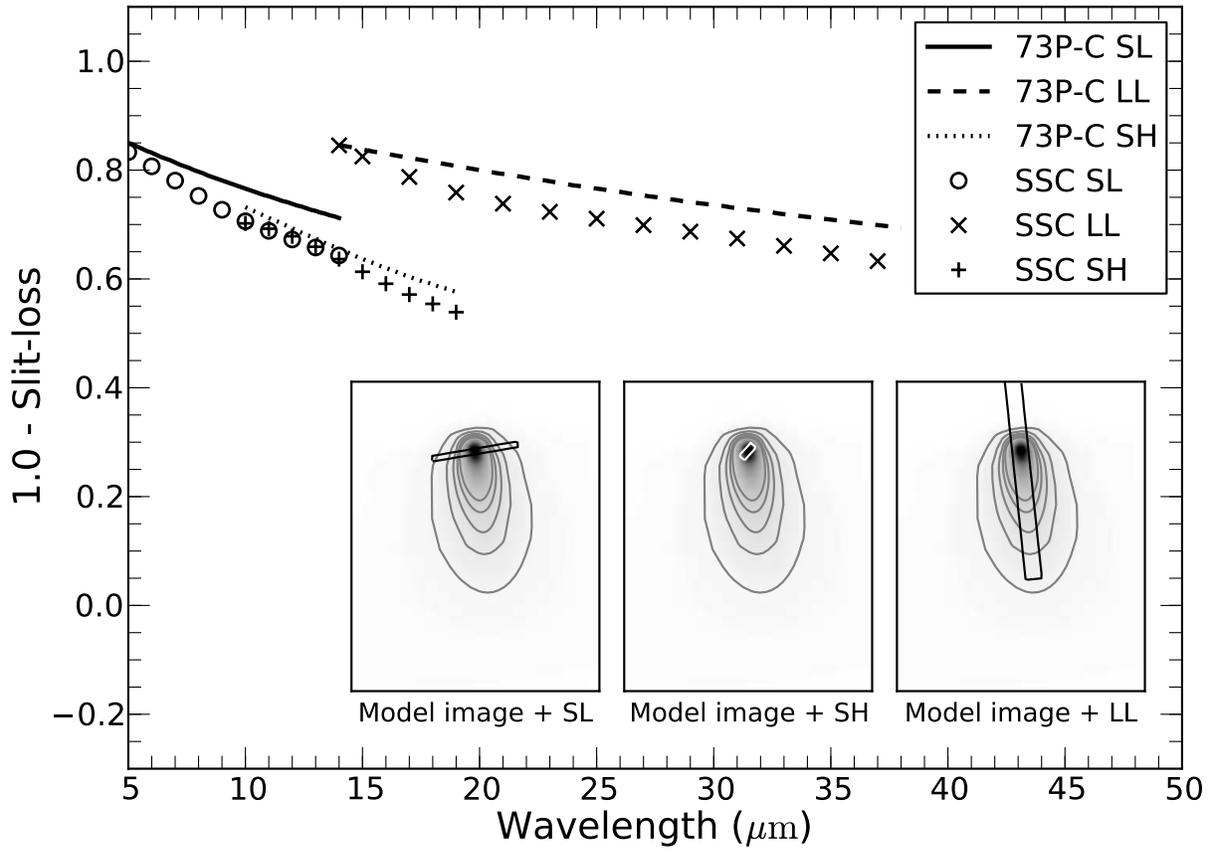}
\caption{ The same as Figure 2, except for fragment C.  The
high-resolution 10--20~\micron{} (SH) module observation is also
included. 
\label{fig:slcfC}}

\end{figure}

We measured the slit-losses of the low resolution modules and the short
wavelength, high resolution module by aligning a rectangular aperture
on each comet's nucleus, reproducing the orientation and size of each
slit with respect to the comet.  The flux contained within the slit is
computed for A) the model image, and B) the model image convolved with
a synthetic 300~K blackbody PSF computed for the IRS instrument at
wavelength $\lambda_i$ and smoothed with a 1.1 (SL) or 1.3 (LL) pixel
wide boxcar function.  Without the smoothing, a Tiny Tim 3000~K PSF
would not match the observed width of calibration stars processed by
CUBISM.  The slit-loss is then $1 - B(\lambda_i) / A$.  Figs.~\ref{fig:slcfB}~\&~\ref{fig:slcfC}  
presents our model images, the orientations of the IRS slits, our
computed slit-losses, and the slit-losses for a point source as
provided by the \textit{Spitzer} Science Center in the \textit{Spitzer} IRS Custom Extraction (SPICE)\footnote{http://ssc.spitzer.caltech.edu/dataanalysistools/tools/spice/spiceusersguide/} software's
calibration files.  To produce the final flux calibrated spectra, we take the extended source
calibrated spectra from SPICE (which assumes an infinitely extended
surface brightness) and divide it by $B(\lambda_i) / A$.  We followed this prescription for all but the SL observations of Fragment SW3-B, because the
tail  of fragment B is smoothly varying over many slitwidths, and the SL slit is
nearly aligned perpendicular to the tail, we elected to
use the SPICE extended source calibration for the SL spectrum of this
fragment.

\begin{figure}
\center
\plotone{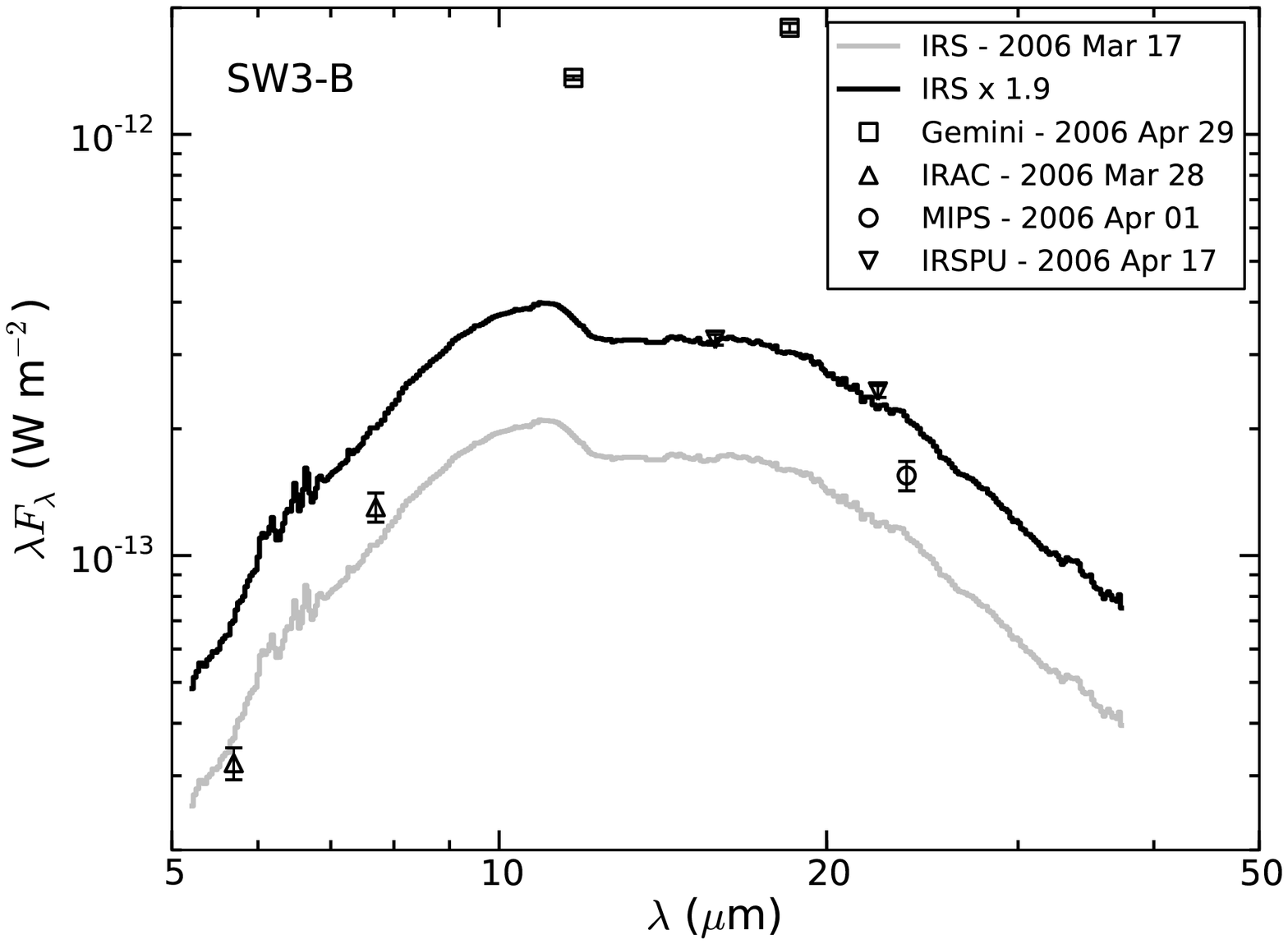}
\caption{The IRS spectrum of 73P/SW3-B. The lower (gray) line is the spectrum
normalized to the flux extracted from the short wavelength slits.  The
upper (black) line is the IRS spectrum scaled by a factor of 1.9 to
match the extracted flux from the 16~\micron{} IRS imaging.  The
Gemini data is from \citet{har10}. The \textit{Spitzer} IRAC and MIPS
photometry is from \citet{reach09}.  The photometry points have all
been scaled to the same effective aperture size ($\rho=2900$ km, or 5
peak-up array pixels), assuming a $\rho^{-1}$ coma profile.  \label{fig:B_irs}}
\end{figure}

\begin{figure}
\center
\plotone{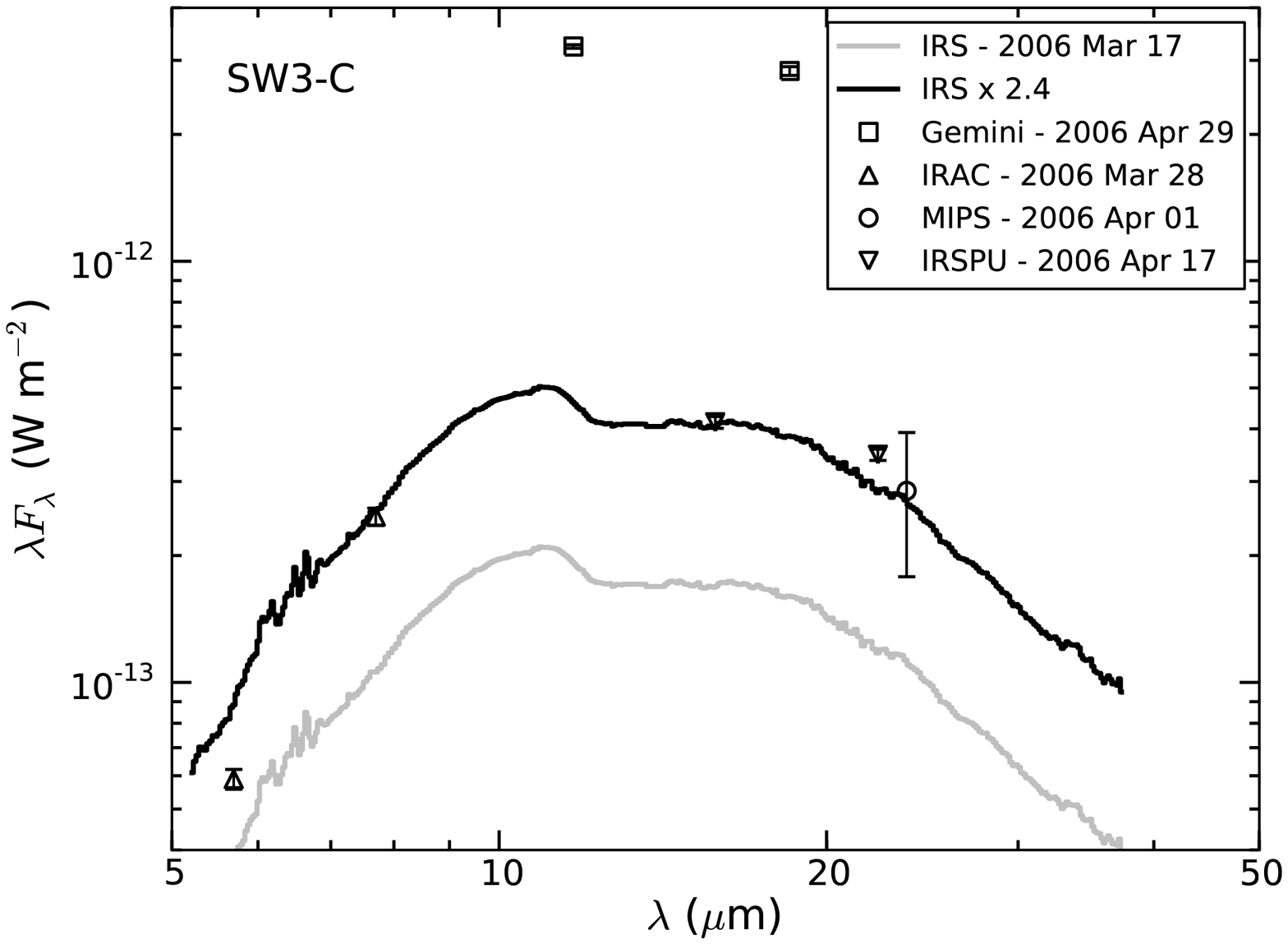}
\caption{The IRS spectrum of 73P/SW3-C. The lower (gray) line is the spectrum
normalized to the flux extracted from the short wavelength slits.  The
upper (black) line is the IRS spectrum scaled by a factor of 2.4 to
match the extracted flux from the 16~\micron{} IRS imaging.  The
Gemini data is from \citet{har10}. The \textit{Spitzer} IRAC and MIPS
photometry is from \citet{reach09}.  The photometry points have all
been scaled to the same effective aperture size ($\rho=5200$ km, or 5
peak-up array pixels), assuming a $\rho^{-1}$ coma profile.  \label{fig:C_irs}}
\end{figure}

The resulting spectra of SW3-B and SW3-C are shown in Figs.~\ref{fig:B_irs}~\&~\ref{fig:C_irs}. For both objects the flux levels for the LL2 (14.0-21.3~\micron{})  and SL2 (5.2-8.7~\micron{}) modes of the IRS have been normalized to that of SL1 (7.4-14.5~\micron{}) at the wavelengths where they overlap, while LL1 (19.5-38.0~\micron{}) has been normalized to the re-scaled flux of LL2. The overall difference in spectral shape between SW3-B and SW3-C occurs because they were observed at different heliocentric distances (1.19 AU for B and 1.47 AU for C) with different local insolation flux densities, and because the particle size distributions for the dust flowing out into the coma were different as well for the two comets - the B-fragment coma contained many more large particles, presumably due to the processes creating the large fragments streaming down the tail by \textit{Hubble} \citep{hal08}. Both result in producing somewhat different equilibrium temperatures for the grains.

\section{The Dust Content of Schwassmann-Wachmann 3}

\subsection{Spectral Modeling}

The infrared emission from a collection of dust near a star is given by

\begin{equation}
F_{\lambda,mod}=\frac{1}{\Delta^{2}}\sum_{i}\int_{0}^{\infty}B_{\lambda}(T_{i}(a,r_{h}))Q_{abs,i}(a,\lambda)\pi a^{2}\frac{dn_{i}(r_{h})}{da}da
\end{equation}

\noindent where $T$ is the temperature for a particle of radius $a$ and composition $i$ at heliocentric distance $r_{h}$, at a distance $\Delta$ from the observer, $B_{\lambda}$ is the blackbody radiance at wavelength $\lambda$, $Q_{abs}$ is the emissivity (emission efficiency) of the particle of composition $i$ at wavelength $\lambda$, $dn/da$ is the differential particle size distribution (PSD) of the emitted dust, and the sum is over all species of material and all sizes of particles for the dust. Spectral analysis consists of calculating the emission flux for a model collection of dust, and comparing the calculated flux to the observed flux. The emitted flux depends on the composition (location of spectral features), particle size (feature to continuum contrast), and the particle temperature (relative strength of short vs. long wavelength features). 

\subsubsection{Scattered Light and Nucleus Contamination}

Prior to the modeling, the effects due to reflected sunlight and nucleus thermal emission need to be checked, and if necessary, removed.   Using photometry obtained on 28 March with the \textit{Spitzer} Infrared Array Camera (IRAC)  \citet{reach09} found the scattered light to be negligible at the 6\% level at 4.5~\micron{} and dropping rapidly at longer (i.e. IRS) wavelengths; the same study also provided an accurate check of the F$_{\nu}$(23.7~\micron{})/F$_{\nu}$ (7.7~microns{}) in our extracted IRS spectra. 

We independently checked the degree to which scattered light might affect the shortest wavelengths of the \textit{Spitzer} IRS (5~\micron{}) of component C with ground-based spectrophotometry covering 0.44-13.5~\micron{} obtained 19 May using The Aerospace Corporation's Broad-band Array Spectrograph System (BASS) and its CCD guide camera on NASA's Infrared Telescope Facility (IRTF), and found it to be less than 11\% and likely smaller (see Appendix). Because the main contributor to the 5~\micron{} emission will be amorphous carbon (see below) its abundance might be too high by 6\%-10\%. Other species will be affected to a lesser extent.

 \citet{har10}, using 10~\micron{} imaging obtained with the Michelle instrument on the Frederick C. Gillett Gemini North (hereafter Gemini North) telescope determined the contribution of the nuclei in a 0.6'' x 1.0'' synthetic beam, and found it to be 1.4\%$\pm$0.2\% for B and no detectable contribution to C (both at a 99\% confidence level). Our \textit{Spitzer} spectra were obtained with a much wider slit and at over twice the distance to the comet, implying an even larger beam/slit dilution factor for an unresolved point source factor, and so we applied no correction for the nuclear emission in this work.

\subsubsection{Composition}

The spectral model used was the same one applied to the \textit{DI} spectra of T1 and the
\textit{Infrared Space Telescope (ISO)} Short Wavelength Spectrograph (SWS) observations of Hale-Bopp (hereafter HB)
by \citet{lisse07}. To determine the mineral composition the observed infrared emission is compared with a linear sum of laboratory thermal infrared emission spectra. As-measured emission spectra of randomly oriented 1~\micron{}-sized powders (produced by grinding of representative mineral samples) are utilized to determine $Q_{abs}$ for a statistical collection of lowest free-energy dust particles. The list of species tested, motivated by their reported presence in interplanetary dust particles, meteorites, \textit{in situ} comet measurements, young stellar objects, and debris disks \citep{lisse07}  includes ferromagnesian silicates of various compositions (forsterite, fayalite, clino- \& ortho-enstatite, augite, anorthite, bronzite, diopside, \& ferrosilite); silicas (quartz, cristobalite, trydimite); phyllosilicates (such as saponite, serpentine, smectite, montmorillonite, \& chlorite); sulfates (such as gypsum, ferrosulfate, \& magnesium sulfate); oxides (including various aluminas, spinels, hibonite, magnetite, \& hematite); Mg/Fe sulfides (including pyrrohtite, troilite, pyrite, \& niningerite); carbonate minerals (including calcite, aragonite, dolomite, magnesite, \& siderite); water ice (clean and with carbon dioxide, carbon monoxide, methane, \& ammonia clathrates); carbon dioxide ice; graphitic and amorphous carbon; and neutral and ionized PAHs \citep{dl07,lisse06,lisse07}. 

\subsubsection{Particle Size and Dust Mass}

Particles of 0.1-2000 ~\micron{} in radius are used in fitting the 5$-$37~\micron{} \textit{Spitzer} data, with particle size effects on the emissivity assumed to vary as 

\begin{equation}
1-Q_{a,\lambda}=[1-Q_{1\mu m,\lambda}]^{(a/1\mu m)}
\end{equation}

The particle size distribution (PSD) is fit at approximately logarithmic steps in radius, i.e., at [0.1, 0.2, 0.5, 1, 2, 5,É100, 200, 500, 1000, 2000]~\micron{}. Particles of the smallest sizes have emission spectra with very sharp features, and little continuum emission; by contrast, particles of the largest sizes are optically thick, and emit only continuum radiation. Spectra with exceptionally strong spectral features  thus have a significant abundance of small ($\sim$ 1 ~\micron{}) grains. 

\begin{figure}
\center
\plotone{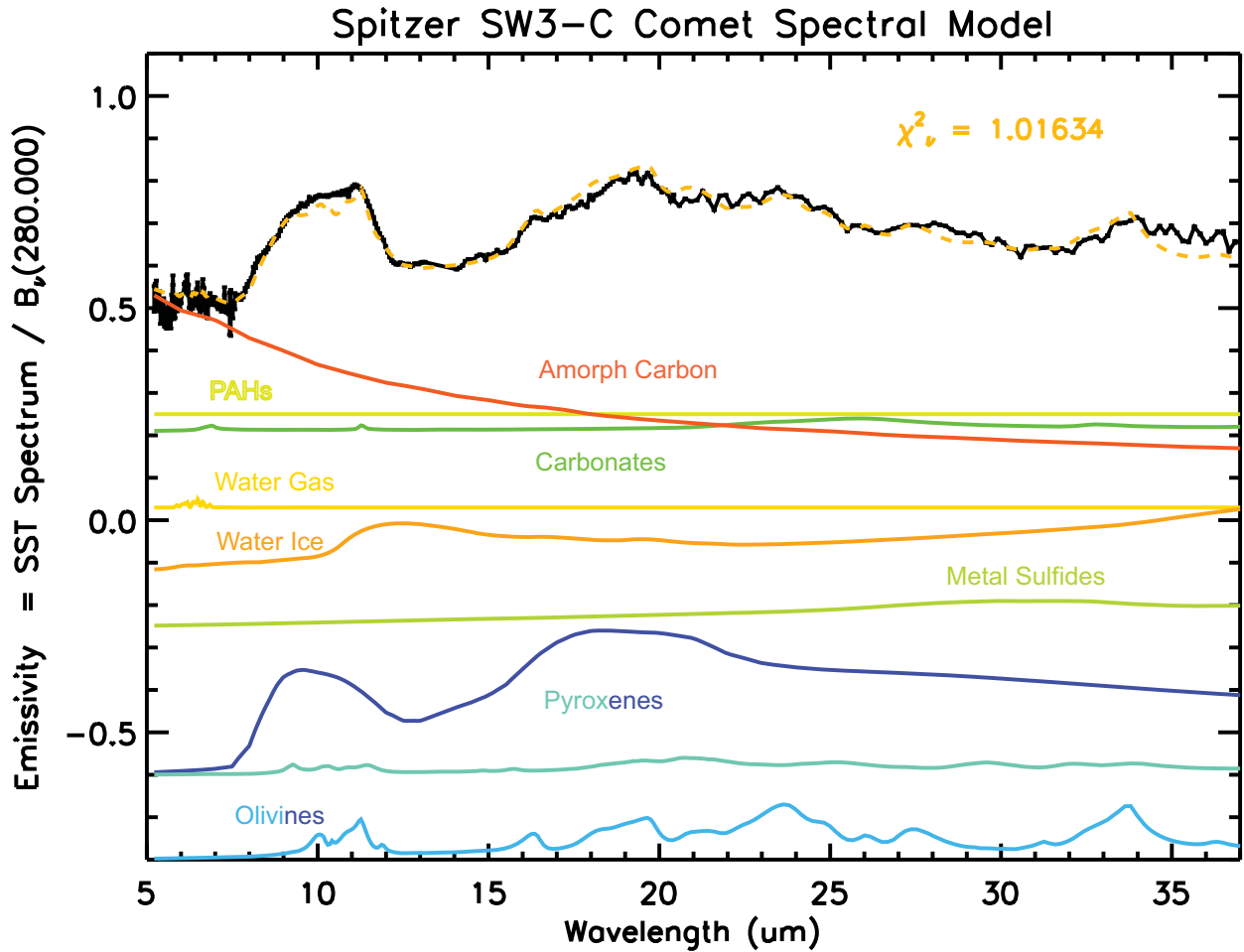}
\caption{The spectral decomposition of SW3-C. In both this component and in B, the material of pyroxene stoichiometry is largely amorphous, while the olivine material is largely crystalline. The shortest IRS wavelengths are dominated by emission from amorphous carbon.\label{fig:Cmod}}
\end{figure}

\begin{figure}
\center
\plotone{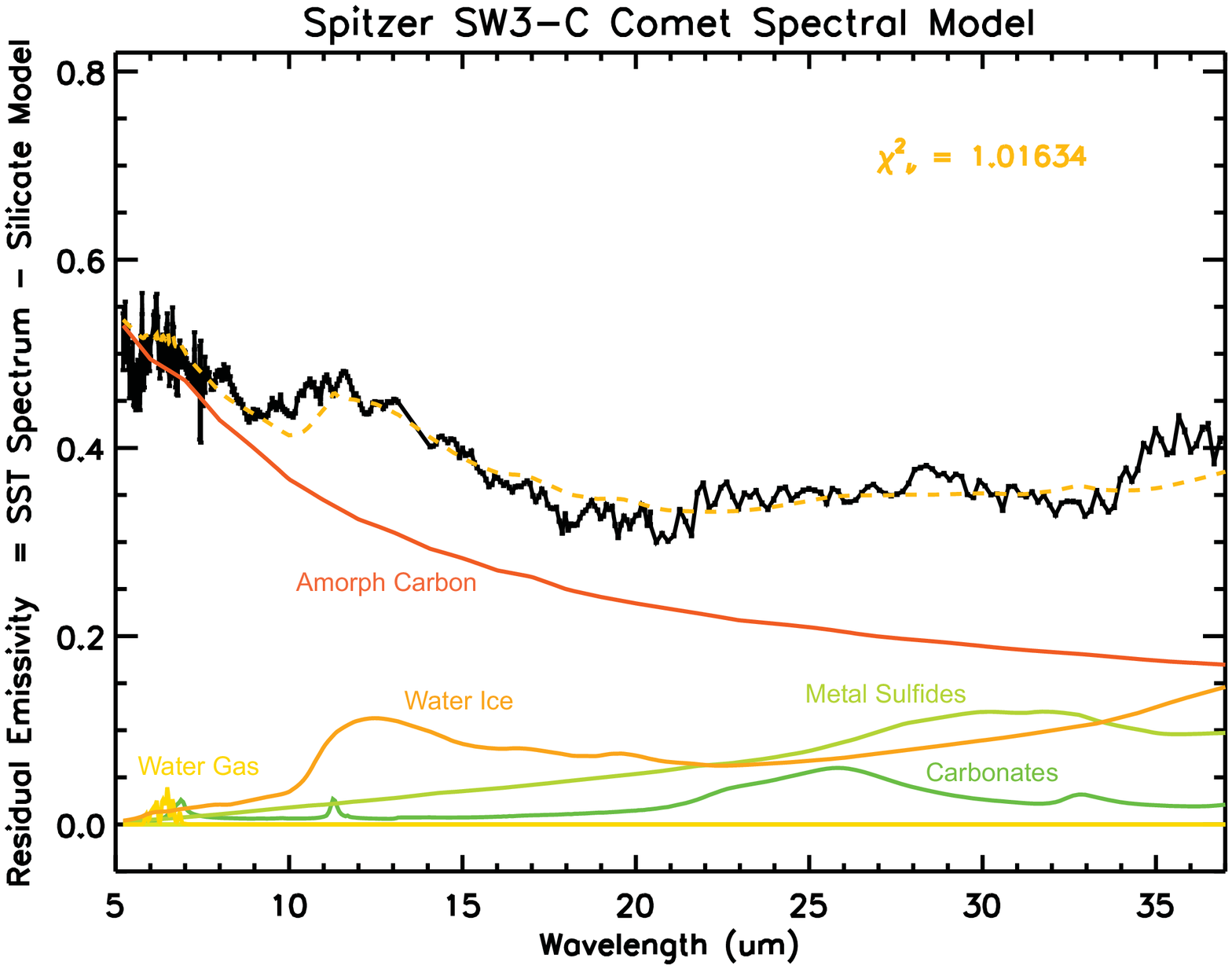}
\caption{The spectral decomposition of SW3-C, after removal of the emission by silicate materials. 
\label{fig:Cmod_detail}}
\end{figure}

As discussed in \S2 above, Fragment 73P/SW3-C was observed at $\Delta$= 0.78 AU, $r_{h}$ =1.47 AU on 17 March 2006. The best-fit PSD found from fitting the IRS spectrum, after division by a 280 K greybody to convert it into an emissivity spectrum (Figs.~\ref{fig:Cmod}~\&~\ref{fig:Cmod_detail}), is a relatively steep one similar to the one found for the very small particle dominated Hale-Bopp coma dust: dn/da = a$^{-4.0}$ (dn/dlog (m) = m$^{-1.0}$, 0.1~\micron{} $<$ a $<$ 1 mm), but with an additional large particle component, a Gaussian (or wide delta function) at $\sim$ 1 mm to fit the observed long wavelength continuum. A similar bump was derived by \citet{vr10} from \textit{Spitzer} images of the comets and their associated meteoriod stream\footnote{A similar 1 mm bump was also present in the PSD of the dust of 1P/Halley \citep{mcdonnell91} and 81P/Wild 2 \citep{tuzzolino04}.}. The total in-beam surface area and mass is 38 $\pm$ 2 km$^{2}$ and 1.6 $\pm$ 0.1 x 10$^{7}$ kg, respectively (equivalent to an 11.5 m radius body of 2.5 g cm$^{-3}$ density). 

\begin{figure}
\center
\plotone{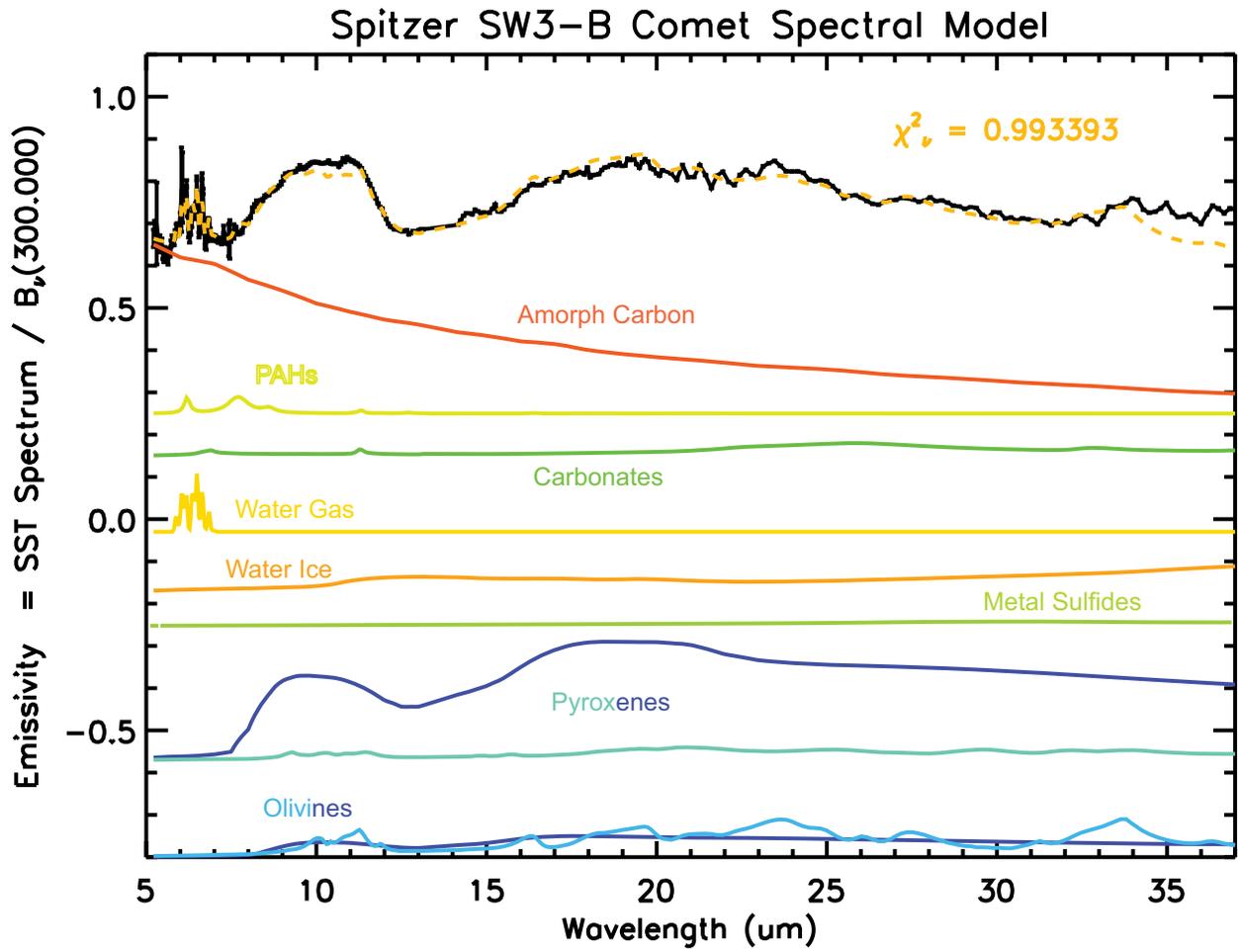}
\caption{The spectral decomposition of SW3-B.  
\label{fig:Bmod}}
\end{figure}

\begin{figure}
\center
\plotone{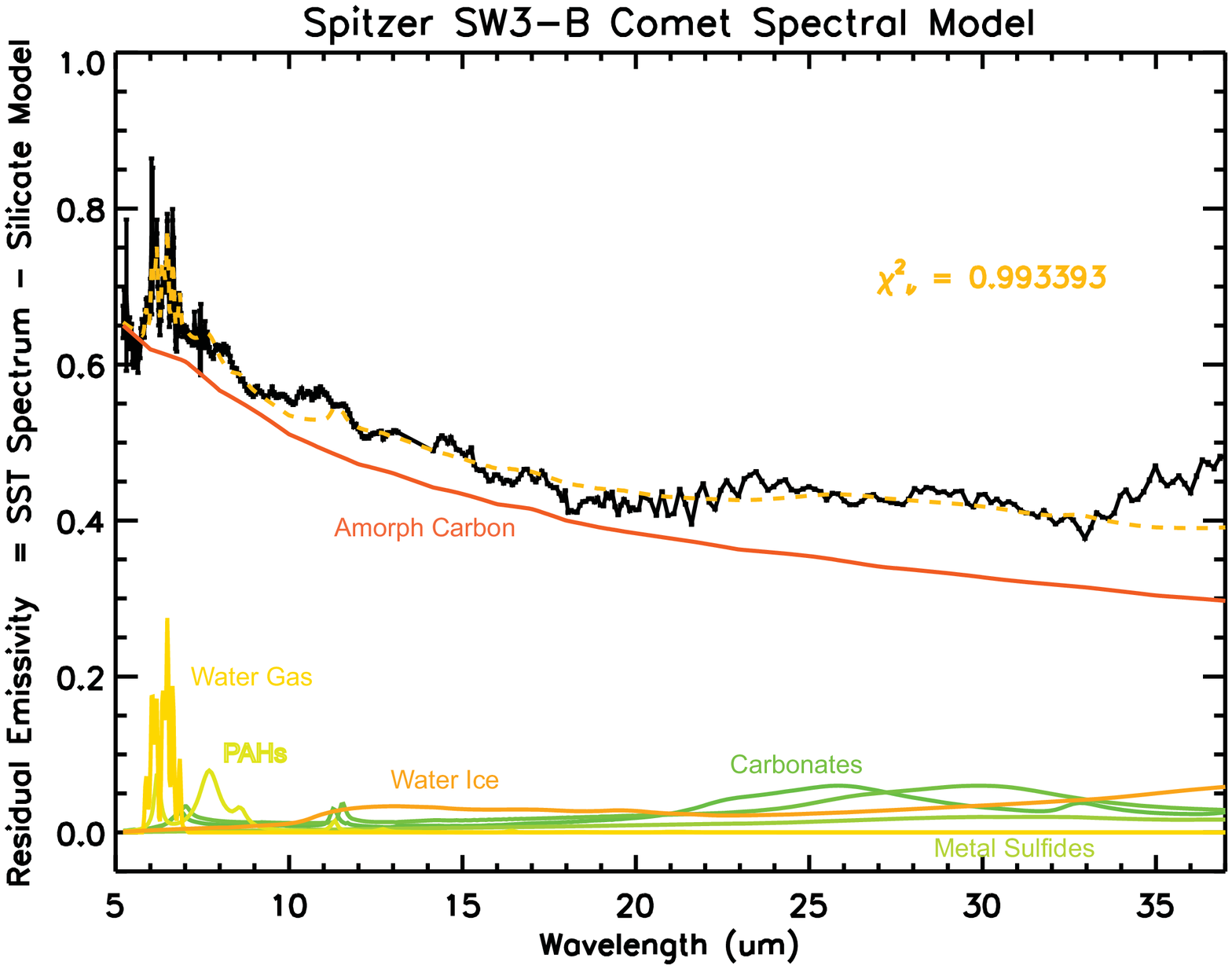}
\caption{The spectral decomposition of SW3-B, after removal of the emission by silicate materials. 
\label{fig:Bmod_detail}}
\end{figure}

\begin{figure}
\center
\plotone{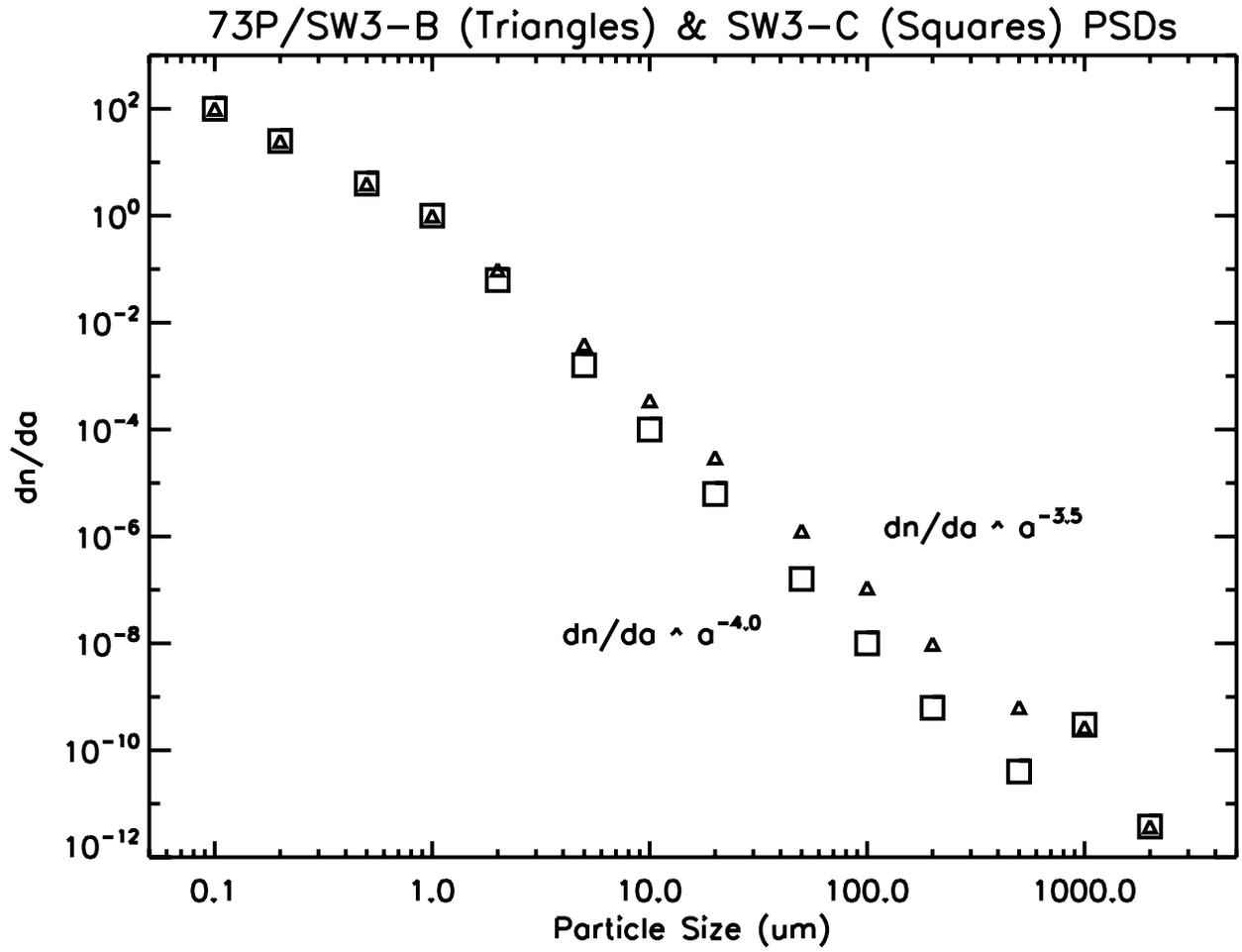} 
\caption{The particle size distributions for the spectral models of SW3. 
\label{fig:psds}}
\end{figure}

Fragment 73P/SW3-B was observed at  $\Delta$ = 0.44 AU,  $r_{h}$ = 1.2 AU on 17-19 Apr 2006\footnote{including all spectral modes}. The best-fit PSD found from fitting the IRS spectrum, after division by a 300 K greybody to convert it into an emissivity spectrum (Figs.~\ref{fig:Bmod} ~\&~\ref{fig:Bmod_detail}), is a moderate one similar to those found for collisional equilibrium  systems: dn/da = a$^{-3.5}$ (dn/dlog (m) = m$^{-0.83}$, 0.1~\micron{} $<$ a $<$ 1 mm), but with an additional large particle component, a Gaussian (or wide delta function) at $\sim$1 mm to fit the observed long wavelength continuum. The total in-beam surface area and mass, dominated by contributions from the largest particles, is 12 $\pm$1 km$^{2}$ and 4.1 $\pm$ 0.2 x 10$^{6}$ kg, respectively (equivalent to an 7.4 m radius body of 2.5 g cm$^{-3}$ density), somewhat less than found in the Fragment C coma, despite the B-Fragment being ~20\% closer to the Sun at the time of observation. The PSDs of both B and C components are shown in Fig.~\ref{fig:psds}.

Extrapolating the in-beam values to the rest of the comae using the peak up images yields a total coma mass of 4-7 x 10$^{8}$ kg and 7-9 x 10$^{8}$  kg for B and C, respectively, contained in the 80 x 56 arcsec field of view of the blue peakup image array. For both fragments, the equivalent single body estimates are probably lower limits. Radar measurements of both fragments by \citet{howell07} indicated the presence of cm-sized ``gravel'' in the comae, which would raise the mass of dust in the comae. For a dn/da $\sim$ a$^{-4.0}$ PSD, we estimate the mass will increase as log(amax), while for dn/da $\sim$ a$^{-3.5}$ PSD, the total coma mass will increase as (amax)$^{0.5}$, where amax is the radius of the maximum sized particle.

The slope of the particle size distributions dn/da = a$^{-4.0}$ and dn/da = a$^{-3.5}$ for C and B respectively, are similar to the value of dn/da = a$^{-3.6~to~4.0}$ found for HB \citep{wil97,lisse99,har02}, which had the most abundant population of small grains measured in any comet.  The high abundance of small particles may explain the unusually high polarization of SW3 compared to that of other ECs \citep{hadamcik09}. A small grain population usually correlates with a strong silicate emission band at 10~\micron{}, and while that band strength in both B and C are stronger than that typically seen in other ECs, they are weaker than that of HB. For fragment C the PSD is at least as steep as that of HB, but a combination of a higher abundance of carbon and a greater fraction of its silicates in the amorphous phase (discussed below),  will result in a more muted silicate band. For fragment B, with its shallower PSD than HB, the inclusion of a large number of ``boulders''  in the spectrograph slit is likely add a significant amount of grain emission with a weak silicate band.

The difference in the power law exponent in the size distributions between the two fragments does not necessarily point to an intrinsic difference in the origin of the grains in the two fragments measured. As discussed above, the profuse shedding of ``boulders/gravel/crumbs'' by B insured a larger collection of bigger particles than for C. Furthermore, there is considerable evidence for ongoing grain fragmentation in the outflowing dust in comets. \citet{tuzzolino04} and \citet{green04,green07} reported that the \textit{Stardust} Flux Monitor measurements of comet Wild 2 contained fine-scale structure that was best understood by particle fragmentation. Similar structure events were detected in the dust outflows of Halley by the \textit{VEGA} instruments \citep{oberc04}. Observations of the material ejected from T1 immediately after the \textit{DI} event are also consistent with extended de-aggregagtion of the ejected material \citep{lisse06,keller07}. The polarimetric imaging data on both large fragments of SW3 itself by  \citet{jones08} indicated grain fragmentation occurred as the dust flowed away from the nucleus.  The radar measurements of \citet{howell07} of SW3 may also be consistent with fragmentation occurring in the coma.  The dust fragmentation will be very stochastic in nature, and the different character of the large-scale fragmentation of the nuclei of B compared to C will further complicate the eventual particle size distributions derived from the spectroscopy.

\subsubsection{Particle Temperature}

Dust particle temperature is determined at the same log steps in radius used to determine the PSD. Particle temperature is function of particle composition and size at a given heliocentric distance. The highest temperature for the smallest particle of each species is free to vary in our model, and is determined by the best-fit to the data; the largest, optically thick particles (2000~\micron{}) are set to the local thermodynamic equilibrium temperature (LTE), and the temperature of particles of intermediate sizes is interpolated at each PSD point between these extremes. Good agreement was found using this method for the temperature of the T1 dust ejecta andthat expected from LTE, where T$_{max}$(best fit) = 1.4 T$_{LTE}$, and by comparison of the T$_{max}$ and T$_{LTE}$ temperatures found for the HB coma dust vs those estimated by \citet{cro97}.

\subsection{Overall Mineralogy}

\begin{figure}
\center
\plotone{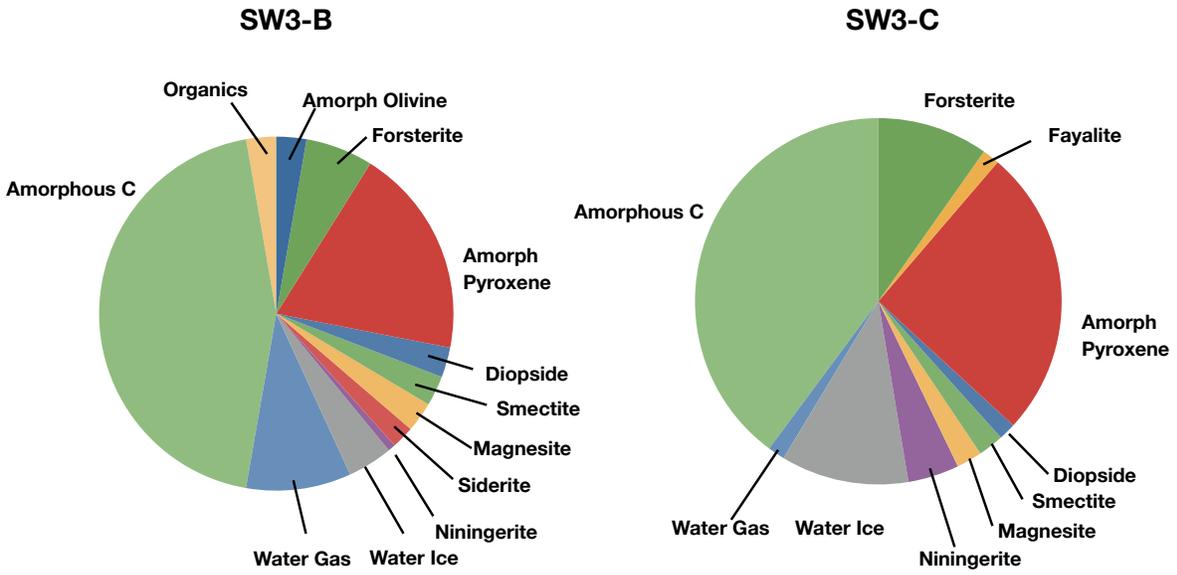} 
\caption{Mineral composition of components B and C. The  individual abundances for both comets are listed in Tables 7 \& 8.
\label{fig:piechart}}
\end{figure}

\begin{figure}
\center
\plotone{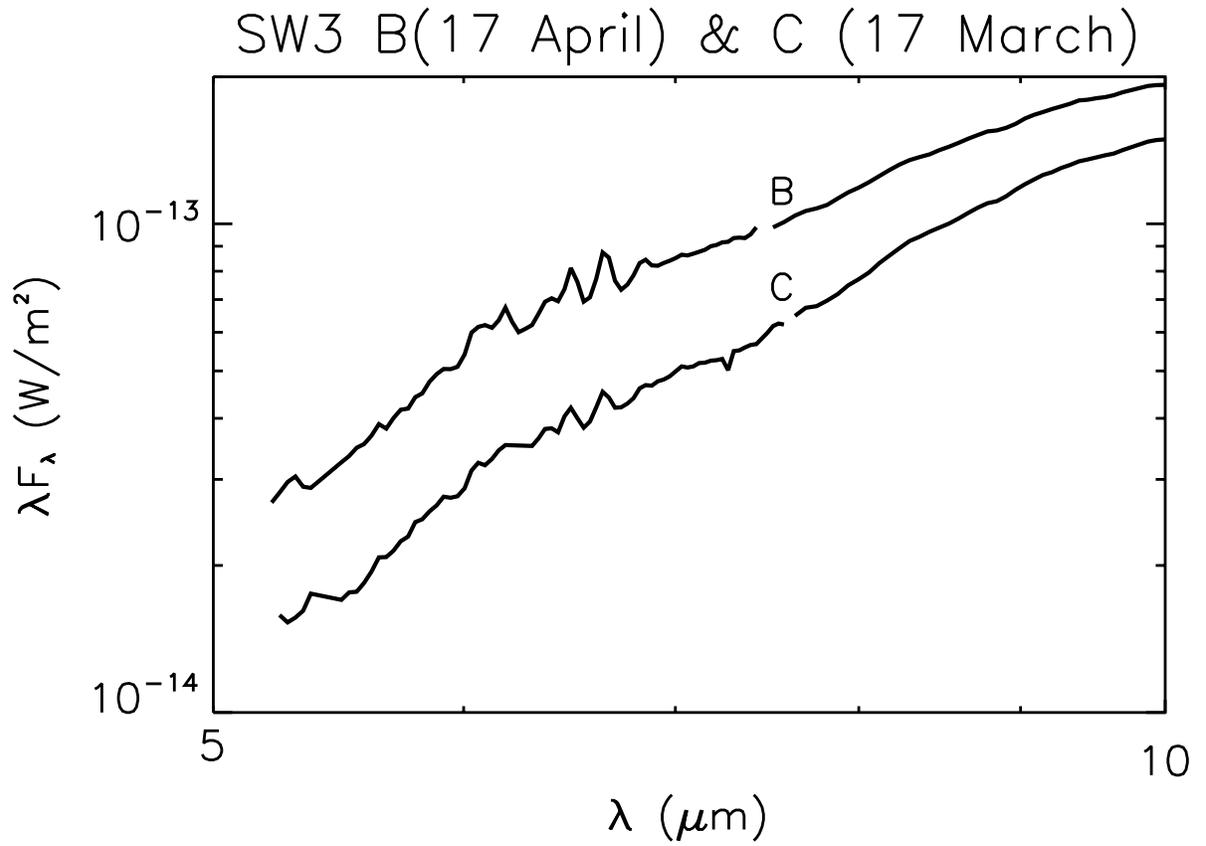} 
\caption{Water vapor bands in B and C compared. As discussed in the text, the strength of water vapor emission was highly time-dependent in both objects, so that any difference in strength was likely simply a matter of the timing of the observations.
\label{fig:water}}
\end{figure}

The overall compositional mineralogy of the dust released from the two largest and brightest fragments of the 2006 apparition, 73P/SW3-C and 73P/SW3-B, is similar, as illustrated in Fig.\ref{fig:piechart}. Both show that is dust dominated by, in order of abundance, amorphous carbon, amorphous silicates of near-pyroxene composition, Mg-rich crystalline forsterite, Fe-metal sulfides, water ice and gas. The major difference between the best-fit models of the two fragments resides in the form of water: in fragment B the effective weighted surface area of water vapor is larger than that of water ice, while this ratio is reversed in fragment C.  The relative strengths of the two fragments' water emission lines above the dust continuum at 6~\micron{} is apparent in Fig.~\ref{fig:water}. PAHs, carbonates and crystalline pyroxenes, found in comets T1, HB, and 17P/Holmes \citep{lisse07,reach10}, are only marginally detected, if at all. The strong similarity between the released dust from the two different fragments is striking, in that they have very different emitted particle size distributions and temporal histories, with the B-fragment going into long extended outburst periods of high activity during April-May 2006, and with the likelihood that the B and C fragments have different vertical structure contributions from the original pre-split nucleus. Nevertheless, the derived \textit{Spitzer} dust compositions are identical within the errors of the modeling, arguing for homogeneity of the original nucleus,  consistent with the narrowband optical photometry and near-IR spectroscopy of gases released from the two bodies \citep{schleicher08,dr07}.

\subsection{Silicates} 

The SW3 silicates are dominated by their amorphous component of near-pyroxene composition, much more so than the dust ejected by the \textit{DI} experiment from T1 or the abundant dust flowing through the coma of HB  \citep{lisse07}. Of the crystalline silicates, only Mg-rich olivine (forsterite) is found in abundance for the best-fit mineralogical models of the dust released from SW3 B \& C.

The presence of crystalline silicates above that seen in the interstellar medium (ISM)
is a common feature of cometary dust. \citet{lisse06}, analyzing the dust ejected
from T1 during the Deep Impact event, found that both olivines and pyroxenes were more
than 70\% crystalline. In HB, the fraction was lower, between 30\% and 60\% \citep{har02,lisse07} In SW3, despite being much more primitive than HB or T1 (Sec.4.2), the dust is still far more crystalline than ISM silicates. The fractional crystallinity of
the silicates in both components of SW3 derived from the Spitzer data is 25\%-30\%. An
independent estimate of the crystallinity of the silicate dust of SW3 by \citet{har10}, using spectroscopic observations of components B and C with the Michelle mid-infrared spectrograph on the Gemini North telescope, and the modeling code of \citet{har02} gives the values of 33.5$^{+25.0}_{-25.3}$\% for B and 25.7$^{+7.2}_{-4.3}$\%.

The 8-13~\micron{} silicate feature strength, ~50\% above the nearby continuum, is significantly higher than the 10-20\% typically seen for ECs, but lower than that seen in HB or the post-impact spectrum of T1. The change in band strength in the pre- and post-impact spectra of T1 is most likely caused by de-aggregation of large, optically thick ($>$ 10~\micron{}) porous dust particles into their sub-micron to micron constituent components  \citep{lisse07}. The same process might account for the strong (compared to other ECs) silicate feature in the SW3 fragments,  and is presumably due to the same mechanism causing the fragmentation of SW3. The even greater strength of the silicate feature in HB and the post-impact spectrum of T1 is likely be related to the more energetic ejection mechanisms involved: the high-energy of the \textit{DI} impactor and the violent outflow from HB compared to the more gentle disruption process in SW3.

\subsection{Amorphous Carbon}

Another striking result of the SW3 compositional modeling is the very large amount of amorphous carbon in the mix. The relative surface area for this component is huge, as can be seen in the very ÒflatÓ emissivity spectrum in the 5-9 $\mu$m range. An independent consistency check of this result was found by applying the Mie dust models of \citet{har04} to the \textit{Spitzer} data, with similar results. There is much more C in the solid phase than for T1 or HB. 

\begin{figure}
\center
\plotone{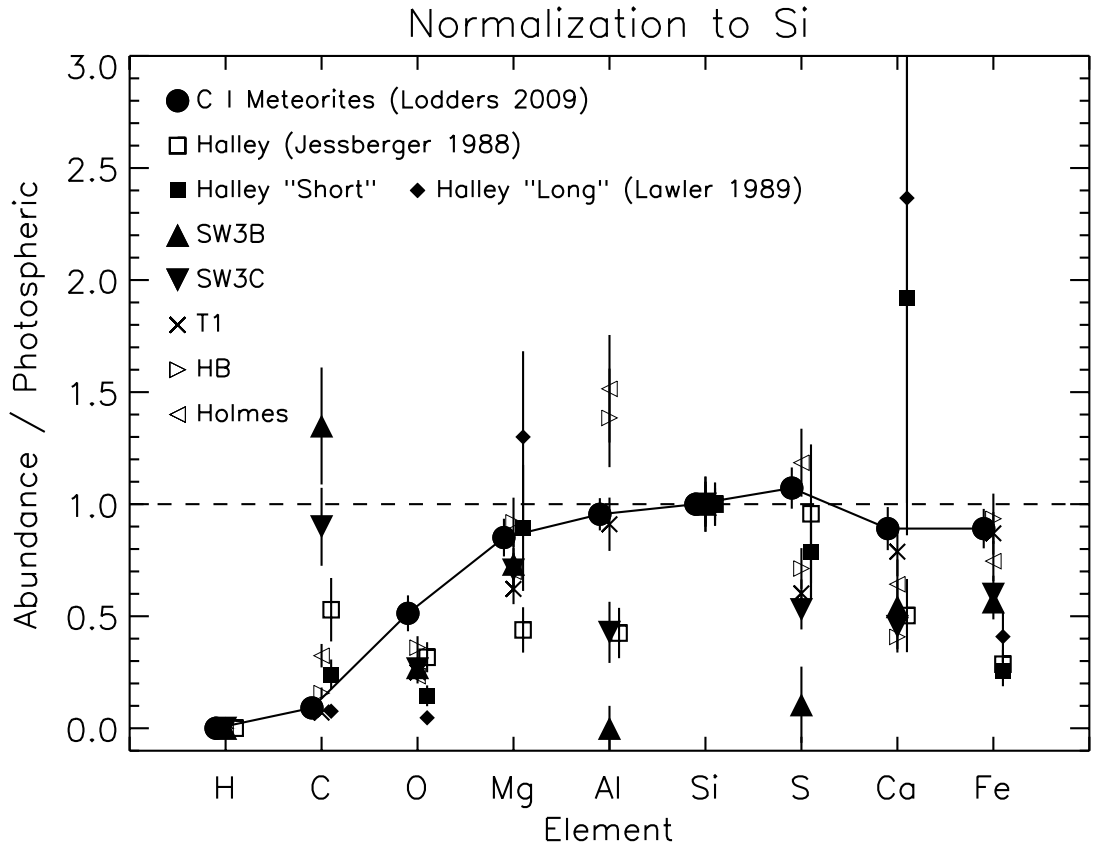} 
\caption{Atomic abundances of SW3 components B and C, compared to solar, normalized to unity for silicon. Also shown are the results for three other comets based on spectral models, the C1 carbonaceous chondrites, and 1P/Halley (based on mass spectrometer results). The solar photospheric  values are taken from \citet{asplund09}, while the CI chondrite values are those of \citet{lodders09a,lodders09b}. The uncertainties for the comet spectra models are \textit{two} standard deviation values . \label{fig:abundances_si}}
\end{figure}

\begin{figure}
\center
\plotone{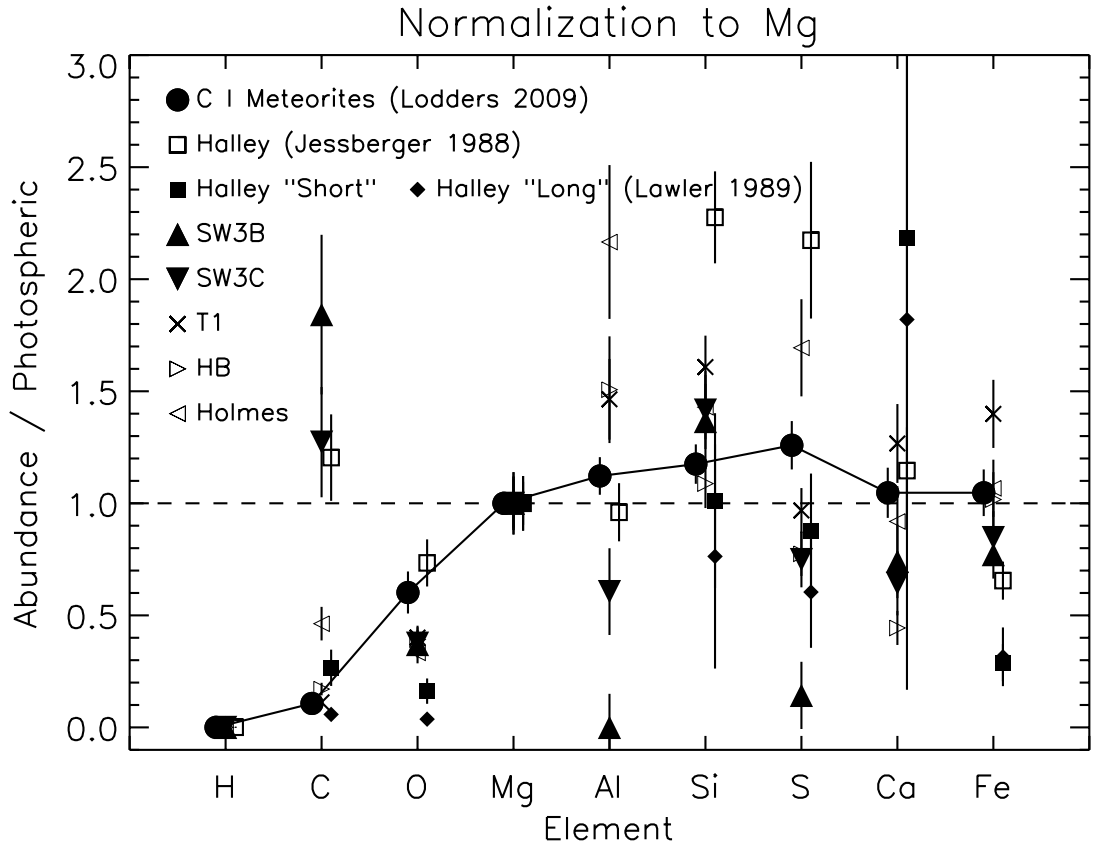} 
\caption{Atomic abundances of SW3 components B and C, compared to solar, normalized to unity for magnesium. \label{fig:abundances_mg}}
\end{figure}

For comparison, the most carbon rich meteorite, Tagish Lake, is 3.6$\pm$0.2 \% \citep{brown00} carbon by mass\footnote{One sample had an abundance of 5.81 \%, higher than any other known chondrite \citep{grady02}, but this is still less than what we derived for SW3.}. Our results for the outflowing SW3 B \& C dust is about an order of magnitude more C-rich by mass.  It must be noted that as the emissivity spectra for amorphous C and nano-phase Fe are both very similar and featureless in the mid-IR, it is near-impossible to spectrally distinguish between the two and thus the total carbon abundance could be somewhat lower  than our estimates. We can, however, put a weak upper limit on the nano-phase Fe content, assuming the elemental abundance of Fe to be less than or equal to solar (Figs.~\ref{fig:abundances_si} and~\ref{fig:abundances_mg}). This implies that nano-phase Fe is able to account for at most $\sim$1/3 of the superabundant amorphous C signal. Additional emission from other spectrally featureless materials, such as very large grains, might also be driving the C abundance high in the \textit{Spitzer} spectra. We note, however, that the \textit{Giotto} and \textit{VEGA} mass spectrometer measurements of Halley have a lower carbon content (compared to the silicate-forming elements Si and Mg) than SW3, and are comparable or slightly higher than T1, HB, or Holmes.

\subsection{Water Ice and Gas} 

Even without application of compositional modeling, the stronger water gas lines in the Fragment B spectrum are clear in Fig.~\ref{fig:piechart}. This finding is somewhat surprising, given the HST imaging of multiple icy bodies moving down the tail of Fragment B \citep{hal06,hal08}, which would seem to imply abundant amounts of solid water ice. On the other hand, the distance from the Sun, 1.2-1.4 AU, and the gross overall best fit dust temperatures, 280-300 K, both suggest that the majority of emitted water should be in the gas phase, and it is possible we are seeing the result of an outburst of water ice some hours or days after it has had a chance to warm up and evaporate. 

Taking note of the much finer size distribution (dn/da $\sim$ a$^{-4.0}$) of emitted dust for Fragment C, its identification as the nominal main nucleus \citep{lamy04,sekanina05}, providing all the other fragments, and its near-normal emission behavior (in terms of the temporal development of its light curve, and the morphology of its coma), the most plausible explanation we can find is that the water emitted from a well mantled (due to fallback of heavy dust fragments) Fragment C is supplied predominantly from subsurface fine ice grains which are slowly subliming as they fly, while the water coming out of Fragment B is flowing out mainly as water vapor, either from the fresh surface of the main fragment B body or the large optically thick (i.e. spectrally undetected) sub-fragments moving down its tail. Evidence for abundant icy dust in the fragment B tail is provided by the narrow ``pinching'' morphological effect seen in the images of B seen in Fig.~\ref{fig:puis} (see also \citet{har10}). Such morphology is predicted by the dynamical models of \citet{lien96}, where the angular width of the observed tail is substantially and atypically narrower than that of the coma, due to ongoing ice sublimation.

Finally, these interpretations are complicated by the fact that the water gas production rate of both components was variable with time. \citet{dr07} observed a decrease in the water production of B by a factor of 6 over a 6-day span of time, and a factor of 2 in a single day, 14.6-15.6 May 2006. The major ``crumbling'' event in B likely occurred near 1 April, closer in time to the \textit{Spitzer} observations, which were likely even more affected by changes in water ice and vapor production. Their data also indicate variability in C, with a 50\% increase (4.8$\sigma$) in a single day. In fact, although the water production rate was three times greater in B than C on 9 April, B's water production rate was half that of C's on 15 April. Thus the relative total strength of the water bands in the two comets is a matter of the timing of the observations, and if the rate at which the emitting species remains in the extraction region of the spectrograph differs for the gas and the dust (which have different trajectories and speeds), the contrast of the water bands over the dust continuum will also simply be a matter of timing.

\section{Discussion}

Comets are composed of interstellar materials that have remained intact, material created during the formation and evolution of the protosolar nebula, and remnant material that was further processed in the nebula at a later time. Due to the presence of a variety of transport mechanisms, material that was once in the innermost regions of the solar nebula may eventually find their way into the comet-formation zone, where they may be incorporated into the comets themselves. The degree to which this occurred, its timing, and the nature of the processing will be imprinted on the comet population. 

The actual chemical content of the material in the early solar nebula is the result of a complex network of gas-gas chemical reactions, gas-grain reactions, dust condensation,  annealing, and and sublimation (see, for example the models of  \citet{gail04} and \citet{wg08}). Because the temperature of the nebula will be highest close to the sun, the processing occurring there is likely to dominate, but imposes the condition that the radial transport must be efficient to introduce the material into the comet-forming region. 

\subsection{Processed Silicates}

The observed degree of crystallinity in SW3 and the other comets studied
spectroscopically is well in excess of that predicted for the first million years by these
models. This suggests that: the process by which crystalline material is produced is more efficient than these
models predict,  the radial mixing is more efficient than predicted, or it lasts much longer than 1 Myr, or other mechanisms for the production and transport of crystalline materials are also in play.

Thermal processing of interstellar dust grains in the inner solar nebula \citep{cc97} offers a natural source for the presence of crystalline silicates in the solar system. The resulting grain chemistry will be dependent on the availability of oxygen. Inside the water dissociation line, OH will also react with C to make CO and CO$_{2}$ \citep{gail01,gail02,wooden08}. Low O fugacity favors the condensation of Mg-rich silicate crystals and Fe metal, and the annealing of amorphous Mg-Fe silicates into Mg-rich silicate crystals (forsterite - Mg$_{2}$SiO$_{4}$, and enstatite - Mg$_{2}$Si$_{2}$O$_{6}$) and Fe metal \citep{gail98,gail04,wooden08}. However, an influx of water from the outer disk can shift the balance toward the production of more Fe-rich crystalline silicates (fayalite -  Fe$_{2}$SiO$_{4}$, and ferrosilite - Fe$_{2}$Si$_{2}$O$_{6}$), and less Fe metal.

While the grain processing is occurring, radial mixing in the protosolar nebula can distribute this processed material beyond 5 AU, into the region where comets were assembled \citep{gail04,kg04,wg08}. However, the balance between crystalline and amorphous material will have a strong radial dependence, as will the mineral content of the material. In the innermost edge of the surviving dust disk, crystalline forsterite dominates, but is quickly replaced by enstatite.  Beyond 10 AU, these pure crystalline species make up only a few percent of the silicate material, which is now almost completely amorphous. Early estimates by \citet{gail04} of the resulting mineralogy near 20 AU placed the fractional composition of enstatite at $\sim$13\% and forsterite at $\sim$1\%. These estimates, however, were dependent on having complete chemical equilibrium between enstatite and forsterite, and more realistic modeling was expected to reduce the fraction of enstatite. In more recent models \citep{wg08} these minerals have roughly equal abundances, but neither are present after 1 million years at more than a few \% beyond 10 AU.

\citet{ciesla09} has modeled the transport of processed inner nebula dust by bipolar jets, where the dust decouples from the gas and is deposited onto the surfaces of the outer disk. The grains are them mixed into the disk by turbulent diffusion. The question remains whether the diffusion can mix the grains into the central plane, where comet growth is likely greatest, on a time scale short enough for that material to be included in the \textit{interiors} of the comet nuclei. The tremendous degree of fragmentation  of component B gives us confidence that we are seeing a substantial portion of what was once the interior of the comet parent body, which probably accreted early in the life of the solar nebula, and crystalline material is certainly present, albeit not in the quantities seen in some other comets, such as HB. 

Similarly, radiation pressure can also move material over substantial radial distances. \citet{dejan09} has shown that this process can be quite effective if the radiation from the disk itself is also included. The disk radiations helps to levitate the grains, so that photons from the star cause the grains to glide over its surface to substantial distances. Whether this mechanism is capable of transporting processed material to the comet-forming region depends on the time scales of over which it operates effectively compared to the availability of processed material to be transported, and the time scale of comet formation.

\citet{hd02} have suggested that nebular shocks at 5-10 AU can produce local temperature spikes that are capable of thermally annealing grains, obviating any need for large scale transport. However, their model predicts essentially no crystalline material should be found beyond 10 AU, and as such would be absent from comets formed in the trans-neptunian region, which is the origin of the ECs, such as SW3.

Another possible mechanism for producing crystalline material closer to the comet-formation zone is through heating by exothermic chemical reactions in organic refractory materials in the grains. \citet{tanaka10} suggest that such reactions may be initiated at temperatures of only a few hundred Kelvins, and are sufficiently energetic to produce temperature spikes capable of annealing the grains. These would still need to be transported to regions in the solar nebula where the temperature has remained below 50 K.

To summarize, a number of mechanisms exist that are capable of producing and transporting processed silicate material into the comet-forming zone. But it is unclear if any can produce the amounts of crystalline material seen in comets (even the relatively low crystalline fraction in SW3) actually observed.

\subsection{Elemental Abundances}

One of the aspects of solar system formation that we want to understand is the distribution of elemental abundances in various solar system materials, and how they are related to the bulk composition from which the solar system formed. In general, the ``cosmic'' abundances, which are really only those representative of stars formed at the galactocentric distance of the Sun, and at the present epoch, are derived primarily from spectroscopic analysis of solar photospheric lines, and laboratory analysis of those meteorites thought to be most ``pristine'', least ``processed'', or ``primitive'' - the CI carbonaceous chondrites. It is to be expected that, as one considers materials condensed in regions of the solar system least hospitable to the survival of the more volatile species, that those atoms significantly locked up in volatile materials would be depleted compared to the Sun,  having been vaporized, and indeed many of them are. \citet{asplund09} summarizes the state of our current knowledge of the abundances of both the Sun and the CI chondrites. Of particular significance to our discussion here are the abundances of C and O, which are major atomic species in potentially volatile materials, and Si, the element to which meteoritic abundances and some cometary abundances are normalized to \citep{asplund09}.\footnote{Results of the impact ionization mass spectrometer data on 1P/Halley have been presented in terms of normalizations to Si \citep{langevin87,lawler89} and to Mg \citep{jck88,jk91}. The use of isotopic ratios to identify ion abundances was stated to be better for Mg by  \citet{jck88}, so we have plotted both. See the Appendix for a more complete discussion.}

Over the years, abundance determinations have evolved, particularly the photospheric ones, due to the difficulty of deriving true abundances from spectroscopic line analysis. For Si, the abundance is currently known to $\pm$0.04 dex, and has not changed by more than $\sim$0.02 dex over the past decade or so. By contrast, the abundances of C and O have changed significantly since the classic work of \citet{anders89}.  Much of the revision was driven by a problem dating back decades, when the depletion of elements from the interstellar gas was being addressed.

The first major analyses of the atomic abundances in diffuse interstellar clouds that included species most easily studied with ultraviolet transitions that became available with the launch of the \textit{Copernicus} observatory. The result was that most species seemed to be depleted relative to solar values (see the reviews of \citet{spitzer75} and \citet{savage96}). In some cases, such as Al and Ca, the depletions in the gas phase with respect to solar were a factor of $\sim$10$^{3}$, while in others, such as C and O, it was between 3 and 5 times. As the measurements became more refined and more lines of sight were sampled, this pattern remained generally the same, with slightly different depletions along different lines of sight. The depletion seemed to be correlated with the condensation temperatures of the main sinks of the elements under circumstellar conditions originally, but later analyses were done within the framework of dense outflows from evolved stars, from which the grains originally condensed (see for example \citet{savage92}). Throughout, these abundances were compared to the most current solar values. Despite the relatively small depletions of C and O in these diffuse clouds, they were important because, due to their overall large cosmic abundances and importance in grain composition, any deficit or excess in the depletions could impact the availability of the elements needed to make the correct amount of dust to reproduce the observed interstellar extinction \citep{cardelli96}. 

At the same time it was realized that the photospheric abundances derived for the Sun, such as those of \citet{anders89} and earlier works, were inconsistent with those derived from young stars being born \textit{today}, with the Sun being overabundant in C and O by a factor of $\sim$2 \citep{sofia94,sofia97}. If the ``cosmic'' abundances were smaller, the discrepancy would be minimized, but would also lower the amount of C and O available to make dust.

Since the work of \citet{anders89} the solar photospheric abundances of C and O have declined significantly. For C, the log of the abundance (where H is defined as 12.00) is, depending on the source, 8.43$\pm$0.05 \citep{asplund09}, 8.39$\pm$0.04 \citep{lodders09a,lodders09b}, or 8.50$\pm$0.06 \citep{caffau10}, down from the value of 8.56$\pm$0.04 of \citet{anders89}. For O, it is 8.69$\pm$0.05, \citep{asplund09}, 8.73$\pm$0.04 \citep{lodders09a,lodders09b}, or 8.78$\pm$0.07 \citep{caffau10}, compared to 8.93$\pm$0.0435  \citet{anders89}. Thus the newer values have dropped by about 13\% \citep{caffau10} to 32\% \citep{lodders09a,lodders09b}, and O has dropped 33\% \citep{caffau10} to 47-48\% \citep{asplund09,lodders09a,lodders09b}. 

It is with these revised ``cosmic'' abundances that modern comet and meteoritic results must be compared. For simplicity, we will use the values from \citet{asplund09} for our analysis of SW3.

\subsubsection{\textit{Caveat Emptor}}

Two issues that are easy to ignore in the spectroscopic analyses of comet abundances are the possible presence of major sinks of elements in undetected species, and the unknowing removal of species from the sample. The first can lead to a ``depletion'' that does not exist. The second to assign a real depletion to the wrong cause.

 \citet{tielens05} notes that the abundance of C and O drop further from their values in the diffuse interstellar clouds discussed previously to denser molecular clouds - the sites of star formation, and the starting point of the solar nebula.  It is possible that much of the O becomes locked up in molecular O$_{2}$ which is not easily measured, while the source of the increased deficit for C remains unclear. The increasing depletion of C and O is independent of any solar abundances, and serves as a warning that both C and O may be stored in forms that are not immediately detectable in spectroscopic studies. Species so stored may or may not later become processed into detectable forms.
 
The C and O may also be preferentially removed from the comet-forming regions long before the comets are assemble. CO (and CO$_{2}$) acts  as a sink for both species. As dust settling to the mid-plane of the nebular disk proceeds, it leaves behind uncondensed gases, such as CO, that will not be incorporated into the comets themselves. Recent observations of young protostellar disks with ages of a few Myr are consistent with the gas remaining in a thick flared disk after the dust has largely settled \citep{acke10}. Bodies being assembled there will be deficient in those elements locked into the major gas species - H$_{2}$ and CO. In this case, it is not simply a matter of the volatiles being too warm to condense, but being significantly separated spatially from the refractories. There is therefore no \textit{a priori} reason to expect the C and O abundances in comets to be exceptionally close to solar to begin with. But comparing them to the abundances of other solar system materials has the potential of shedding light on the evolution of the solar nebula.

\subsubsection{Rock-Forming Elements}

Figs.~\ref{fig:abundances_si} and ~\ref{fig:abundances_mg} show the abundances of a number of elements relative to both Si and Mg in SW3 and three other comets studied spectroscopically with either \textit{Spitzer} or the \textit{Infrared Space Observatory (ISO)}, normalized to the solar photospheric abundance ratios, taken from \citet{asplund09}. For the majority of the rock-forming elements, SW3 is similar to the other comets, and close to or somewhat less than solar. For comparison, we also show the abundances for 1P/Halley, based on the impact ionization mass spectrometers of the \textit{Giotto} and \textit{VEGA} spacecraft based on two studies, that of  \citet{jck88} and \cite{jk91}, and that of \citet{lawler89}. Also shown are the mean abundances for the CI carbonaceous chondrites, from \citet{lodders09a,lodders09b} (also listed in \citet{asplund09}). For Mg (in the Si-normalized plot), Si (Mg-normalized), and Fe (both normalizations), the abundances of all five comets studied spectroscopically, including both components of SW3, are consistent with both the CI and solar photospheric values. For Al, S, and Ca there is considerable scatter around the the CI and photospheric values for SW3, but the uncertainties are larger, as these elements are present only in minerals with trace amounts (diopside, smectite nontronite, and niningerite) in the models. The abundances of these elements in these five comets are also consistent with the mass spectrometer data on Halley, although the uncertainties for Ca are large in \citet{lawler89}, and they did not report an abundance for Al.

\subsubsection{The Oxygen Abundance}

For all of the objects shown - the five comets studied spectroscopically, both sample results on Halley, and the CI chondrites, O is depleted with respect to the solar values.  The spectral studies specifically include H$_{2}$O vapor and ice as components, while the Halley data do not, and the chondrites are not as water-rich as the comets. The depletion of O in the comets determined from the spectral models might result from:

$\bullet$ the water model underestimating the actual amount of H$_{2}$O \textit{vapor} present

$\bullet$ the spectral model underestimating the H$_{2}$O \textit{ice} abundance

$\bullet$ the dynamics of the gas compared to the entrained dust causing the H$_{2}$O to be underestimated compared to the dust (Si, Mg, etc.)

$\bullet$ O being sequestered in some dust species to which the spectral models are insensitive (large or spectrally featureless grain materials)

$\bullet$ O being in gaseous species not detected in the spectra

$\bullet$ O being truly depleted in comets

The H$_{2}$O vapor abundances are determined by the bands in the 6 $\mu$m region, which are well-defined, and would need to be off by a factor of $\sim$3 (regardless of which normalization is used) to bring them into agreement with solar values. 

The H$_{2}$O ice band occurs at the minimum between the 10 $\mu$m and 20 $\mu$m silicate bands and is not easily masked or blended with other features. Very large ice particles that were missed in the particle size distribution would be spectrally weak and might go undetected. The likely presence of large chunks of ice in component B might explain its lower spectrally-detected abundance than in C. But C is also under-abundant in O, so this cannot be the sole answer either.

For a fixed spectral extraction size, faster-moving species, such as gas molecules,  will travel out of the extraction zone than the more slowly-moving dust grains.  For a grain ejection speed imparted by the outflowing gas,

\begin{equation}
v_{ej}\sim2.3(\rho a_{cm})^{-1/2} m s^{-1}
\end{equation}

\noindent \citep{reach10} a grain with radius $a\sim$ 1 $\mu$m and $\rho\sim$ 2 g cm$^{_3}$ will be moving with $v \sim$ 0.3 km s$^{-1}$, or about 1/3 the speed of the gas. So compared to the dust grains responsible for the spectral features modeled, the water gas will be under-represented, which would cause O to be under-represented. Note that this also would suggest that larger spectrally gray grains may be over-represented in the extraction aperture, as they lag the smaller grains. This may influence the PSD derived from the spectral fitting, which is further complicated by the presence of multiple ``nuclei'' in the extraction aperture for fragment B.

The \textit{Stardust}  sample of comet Wild 2 contains dust material that does not fall into the minerals used in the spectral fitting. The largest grain first examined in detail was one similar to calcium-aluminum-rich inclusions (CAIs) found in meteorites \citep{zol06}, which contain many oxygen-bearing minerals. The \textit{Stardust}  grain contains spinel (MgAl$_{2}$O$_{4}$), anorthite (CaAl$_{2}$Si$_{2}$O$_{8}$ in its pure form), and gehlenite (Ca$_{2}$Al[AlSiO$_{7}$]). Other meteoritic CAIs are generally rich in SiO$_{2}$, Al$_{2}$O$_{3}$, MgO, and CaO groups \citep{krot01}. Such mineral species, especially if they were in 1 mm sized grains, would be missed in the spectral modeling. A comparison of the \textit{Giotto} and \textit{VEGA} probes showed that the scattered light was dominated by mm-sized grains \citep{mcdonnell91}. In SW3 itself, analysis of the \textit{Spitzer} Miltiband Infrared Photometer (MIPS) images of the dust tails and dust trails indicates that the ``bump'' seen in the PSD continues to cm-sized grains \citep{vr10}. This is supported by the radar measurements \citep{howell07}. So the contribution by large grains to the O abundance remains uncertain, but may be significant.

In principle, CO and CO$_{2}$ gas might be another possible sink for the O not detected in the spectral modeling. CO is known to be severely depleted in SW3 in particular, compared to other comets \citep{disanti07}. If CO is a major source of O-depletion from the dust, it is not in the form of \textit{cometary} CO in SW3. There is tremendous scatter in the abundance of CO amongst the comet population (see below) and much of this may be the result of the separation of the dust and gas in the solar nebula, similar to that seen in young protostellar disks, as discussed previously.

Another source of the depletion of O in the spectral studies is its incorporation into organic species. In the Halley samples, the bulk of the C is in for form of complex organics - the ``CHON'' material \citep{fomenkova92,fomenkova94,fomenkova99}. For the 40 CHON particles studied by \citet{jck88}, the ratio of O to C atoms for the sample has considerable scatter ranging from 0.2 to 10. In fact, \citet{jk91} concluded that ``most of the oxygen is not from the silicates, but from an organic phase.'' Doubling the O abundance in the spectrally analyzed comets would bring their values of O closer to that of the CI chondrites, but probably not to solar values. The O stored in organics \textit{is} included in the Halley sampling, but not the O stored in H$_{2}$O, as the survival of ice particles at the distance of the Halley sampling is thermodynamically excluded \citep{jck88}. A gas/dust ratio of $\sim$2 would bring the Jessberger et al. abundance close to solar, while the ratio would have to be considerably larger, $\sim$5 or more, for the abundance determination of Lawler et al. Thus both the \textit{in situ} sampling and spectral models of comets may be missing significant O-bearing material, but a different material in each case.

In summary:

$\bullet$ O seems to be depleted in all comets, including SW3, compared to solar values.

$\bullet$ some of the under-abundance may be due to under-sampling the O tied in H$_{2}$O gas in the spectral models, which exits the sampling beam faster than the dust.

$\bullet$ some of the under-abundance in the spectral models may be due to not including O-rich material that is tied up in large rocky grains and CHON material.

$\bullet$ the \textit{in situ} measurements of Halley, which include some large grains and O-bearing CHON, are still somewhat under-abundant in O, but the O in the form of H$_{2}$O is missed. If the dust-to-gas ratio is one or more, the results approach the solar values.

$\bullet$ A lot of O was likely tied up in CO gas in the nebula which remained ``at altitude'' while the dust settled into the mid-plane of the solar nebula. Thus most of the objects in the solar system may have formed under conditions where H, C, and O were depleted.

\subsubsection{The Carbon Abundance}

One of the most striking aspects of the model fits to the \textit{Spitzer} spectra of SW3 is the high abundance of amorphous carbon. While the C abundance in most comets is sub-solar, and similar to that of carbonaceous chondrites, it appears near to solar in SW3, within the uncertainties, which are somewhat larger than the other objects studied. It is also considerably higher than the Halley analysis by \citet{lawler89} but marginally consistent with that of \citet{jck88} which is at least a factor of two higher than that of Lawler et al.

Using an independent set of observations of SW3 with the Michelle spectrograph on the Gemini North telescope, \citet{har10} modeled both components of SW3 with the \citet{har02} thermal emission code, and found that the the relative abundance of amorphous C in the anti-sunward coma was between $\simeq$ 45-23 wt\% for B and $\simeq$60-42 wt\% for C. Using a similar treatment, the amorphous C abundance was determined to be 21 wt\% for HB, 15 wt\% for C/2001 Q4 (NEAT), and 28\% for T1. The fractional abundance of carbon compared to silicates decreased with increasing distance from the nuclei of both components. The median values within the ranges found from the Michelle data are significantly higher for SW3-B that either HB or T1, while SW3-C was even higher. Thus these independent analyses, using different mineral input spectra, a different modeling code, and observing the SW3 components at a different time than the \textit{Spitzer} observations, yielded similar results for the high abundance of C in SW3 compared to other comets.

At the same time, SW3 is highly deficient in almost all C-bearing molecules except HCN and CO$_2$ \citep{villaneuva06,disanti07,dr07,lis08,reach09}. Carbon combustion in the inner solar nebula is expected to result in the removal of amorphous C and its conversion to CO, CO$_{2}$ and various C-bearing hydrocarbon molecules  such as CH$_{4}$ and C$_{2}$H$_{2}$, although others such as C$_{2}$H$_{6}$ and CH$_{3}$OH require a different source \citep{gail02,wg08}.  CO was detected in SW3-C by  \citet{disanti07} with an  abundance of CO/H$_{2}$O = 0.5\% $\pm$ 0.13\%, making it at the lowest end of of the observed ratio in comets. \citet{dr07}  reported that SW3 was depleted in most carbon-chain molecules, compared to the majority of other comets studied so far. The high abundance of amorphous C and simultaneous low abundances of CO and carbon-chain organics are all consistent with SW3 being formed in a region that did not inherit the products of C-combustion in the inner solar nebula. Further depletion of gaseous C-molecules compared to solid amorphous C will occur as dust sedimentation coours in the disk mid-plane, as was discussed for previously for atomic O. The one discrepancy is CO$_{2}$. \citet{reach09}, using Spitzer imaging data, derived CO$_{2}$/H$_{2}$O ratios of 10\% and 5\% for B and C, respectively, which indicates that CO$_{2}$ was not depleted.

A significant (if not well-quantified) fraction of cometary CO may come from a ``distributed'' coma source that might include refractory material, or from a complex network of chemical (and photo-chemical) reactions \citep{pierce10}, and might behave differently. The CO abundance varies across the comet population by as much as a factor of 40 \citep{mumma03,bm04} and as yet does not seem to be well-correlated with other species. A significant complication arises from the CO band optical thickness. \citet{bm10} have recently analyzed both the CO $J$(1-0) and $J$(2-1) lines in Hale-Bopp, and have concluded that the CO $v$=1-0 rovibrational lines often used to detect CO - and provide evidence for a distributed source - are severely optically thick for heliocentric distances of 1.5 AU or less, and that in fact there is \textit{no} evidence for a distributed CO source in this comet.

One added uncertainty in the determination of the elemental carbon abundance is our uncertain knowledge of the amount of metallic Fe present in the dust, in the form of bulk Fe or FeNi, which could contribute to a featureless continuum in the models, and be responsible for some of the continuum attributed to amorphous C. Metallic Fe is a well-known component of chondritic porous interplanetary dust particles (IDPs), such as the GEMS (\textit{G}lass with \textit{E}mbedded \textit{M}etals and \textit{S}ulfides) \citep{bradley92,bradley94}. \footnote{While the GEMS themselves possess infrared spectral features resembling the amorphous dust of the interstellar medium, the IDPs in which they are embedded are similar to those observed in comets \citep{bradley99}. The presence or absence of GEMS material in the \textit{Stardust}  samples is complicated by the melting of the aerogel in the high-speed capture process.}.

Finally, scattered solar light will contribute most strongly at the shortest wavelengths - precisely where the models are most sensitive to the carbon content. We have placed an upper limit on this contribution at 11\% at 5 $\mu$m (Appendix). This would reduce the amorphous carbon contribution by at most 10\%, not substantially affecting our compositional results.

In summary, the abundance of amorphous carbon derived from the spectral models

$\bullet$  appears equal for B and C, within the uncertainties

$\bullet$ is higher than the majority of other comets studied spectroscopically (in both independent studies)

$\bullet$ is consistent with it a low abundance of CO and carbon-chain organic molecules if SW3 did not incorporate significant amounts of C-combustion products from the inner solar nebula

In terms of the \textit{atomic} abundance (not the form that it is in):

$\bullet$  carbon appears nearly solar in both components of SW3, although the uncertainties are large.

$\bullet$ carbon is somewhat higher in abundance  than other comets whose composition is determined from spectral decomposition, and possibly higher than the \textit{in situ} measurements of Halley.

$\bullet$ the slightly higher atomic C may result from it being sequestered in a moderately refractory \textit{solid} phase, instead of highly volatile ices

\subsection{Comet Homogeneity}

If the mineral content of the region where cometesimals grew and were assembled into their final cometary form occurred in a region where the composition of the dust material was evolving with time due to the delivery of processed material from the inner disk, we would expect that individual cometary nuclei would be chemically and mineralogically heterogeneous. Their surfaces would then have a different mineralogy than their interiors. 

The fragmentation of SW3 exposed material that was once deeply buried in its interior, and the further disintegration of component B assured a greater exposure of this material than in C alone.  Both components have approximately the same carbon content (based on independent observations and spectral models) and degree of silicate crystallinity, suggesting that they formed from the same mix of materials. If B is overall exposing more deeply-stored materials than C, this would suggest that the components were assembled at a time when ``salting'' of the refractories by materials processed in the inner solar nebula was minimal. There is no ``time signature'' evident. 

Given this argument and the compositional similarity of fragments B and C in the more refractory materials, as well as gaseous species \citep{dr07}, we may conclude that there was a substantial homogeneity in the pre-breakup nucleus of SW3.  Furthermore, if we consider that the radius of fragment B is $\leq$0.4~km \citep{howell07}, and fragment C is $\sim$0.7~km \citep{toth05, howell07}, then we may consider that perhaps the entire pre-breakup nucleus is homogeneous throughout. 

\section{Conclusions}

The inner solar nebula possessed the thermal and chemical ingredients for the production of silicate material significantly more crystalline than the raw interstellar dust from which it formed. Gas-gas and gas-grain chemical reactions likely converted some portion of the primordial amorphous carbon in ISM grains into a variety of carbon-containing molecules. Both the crystalline material and the amorphous C-depleted and hydrocarbon molecule-rich material were mixed radially, although cometary material formed very early or far from the inner nebula would not be affected to the degree of those formed later or closer to the sun. Spectroscopic observations of both B and C components of 73P/Schwassmann-Wachmann 3 reveal a low degree of grain crystallinity compare to other comets (but higher than the ISM), higher amorphous carbon content than most other comets, and depletion of most C-rich molecules relative to other comets. These are all indicative of being more primitive than other comets studied remotely, and suggest it was formed earlier than most, or further from the warm inner nebula, or both.

No significant difference in the crystallinity nor amorphous carbon content was observed between the two components in either the \textit{Spitzer} data nor the independently-observed and modeled Michelle spectra by \citet{har10}. The difference in the rates of nucleus fragmentation suggests that we are likely digging deeper into the pre-fragmentation nucleus in B than in C. Despite this, they appear spectroscopically very similar, arguing against significant heterogeneity and/or layering. 

One of the most significant differences observed, the relative strength of the water vapor bands, may not point to any intrinsic difference in the two components, but instead to their being significant temporal changes in their water vapor production, or significant differences in the distribution of water ice sources in the extraction beam.

All comets studied spectroscopically exhibit a general under-abundance of O-rich species. It is likely that much of the O is tied up in grains that are spectroscopically featureless, or at least not easily detectable with current techniques. These do not, however, explain the deficit in the O abundance when measured \textit{in situ} (such as Halley), which is likely dominated by the exclusion of the O on the form of H$_{2}$O gas. Some of it may be in the disk gas that that remains at high altitude in the proto-solar nebula as the dust from which comets form settle to the mid-plane, as has been seen in other  protostellar disks. 

A better understanding of the chemical and physical evolution of the early solar system through the analysis of cometary material will require many studies using a variety of complementary techniques, both remote sensing (spectra, imaging) and \textit{in situ} sampling (dust fluence experiments, sample returns, remote sample analysis).

\acknowledgments

This work is based on observations made with the \textit{Spitzer Space Telescope}, which is operated by the Jet Propulsion Laboratory, California Institute of Technology under a contract with NASA. Support for this work was provided by NASA through an award issued by JPL/Caltech, under JPL contract numbers 1275586, 1274485, 1286245, 1287505. This work was supported in part by the Aerospace CorporationÕs Independent Research and Development (IR\&D) program. C. M. Lisse  gratefully acknowledges support for performing the modeling  described herein from the APL  Janney Fellowship program. MSK acknowledges support from NASA Planetary Astronomy Grant NNX09AF10G. MLS acknowledges support from the AAS Small Research Grant program. The authors would like to thank Vikki Meadows and Nickolaos Mostrodemos for their help in preparing for the \textit{Spitzer} observations, Mike Cushing for his helpful suggestions in processing the SpeX data. The authors would also like to  thank W.T. Reach for many valuable discussions concerning  this work, Xander Tielens for his insight on the depletion of C and O in the molecular cloud environment,  and Sechii Yoshida for use of the comet light curves.

{\it Facilities:} \facility{Spitzer (IRS)}, \facility{IRTF (SpeX,BASS)}

\appendix

\newcommand\afrho{$A(\theta)f\rho$}

\section{Ground-Based Observations of 73P/Schwassmann-Wachmann 3}

\begin{figure}
\center
\plottwo{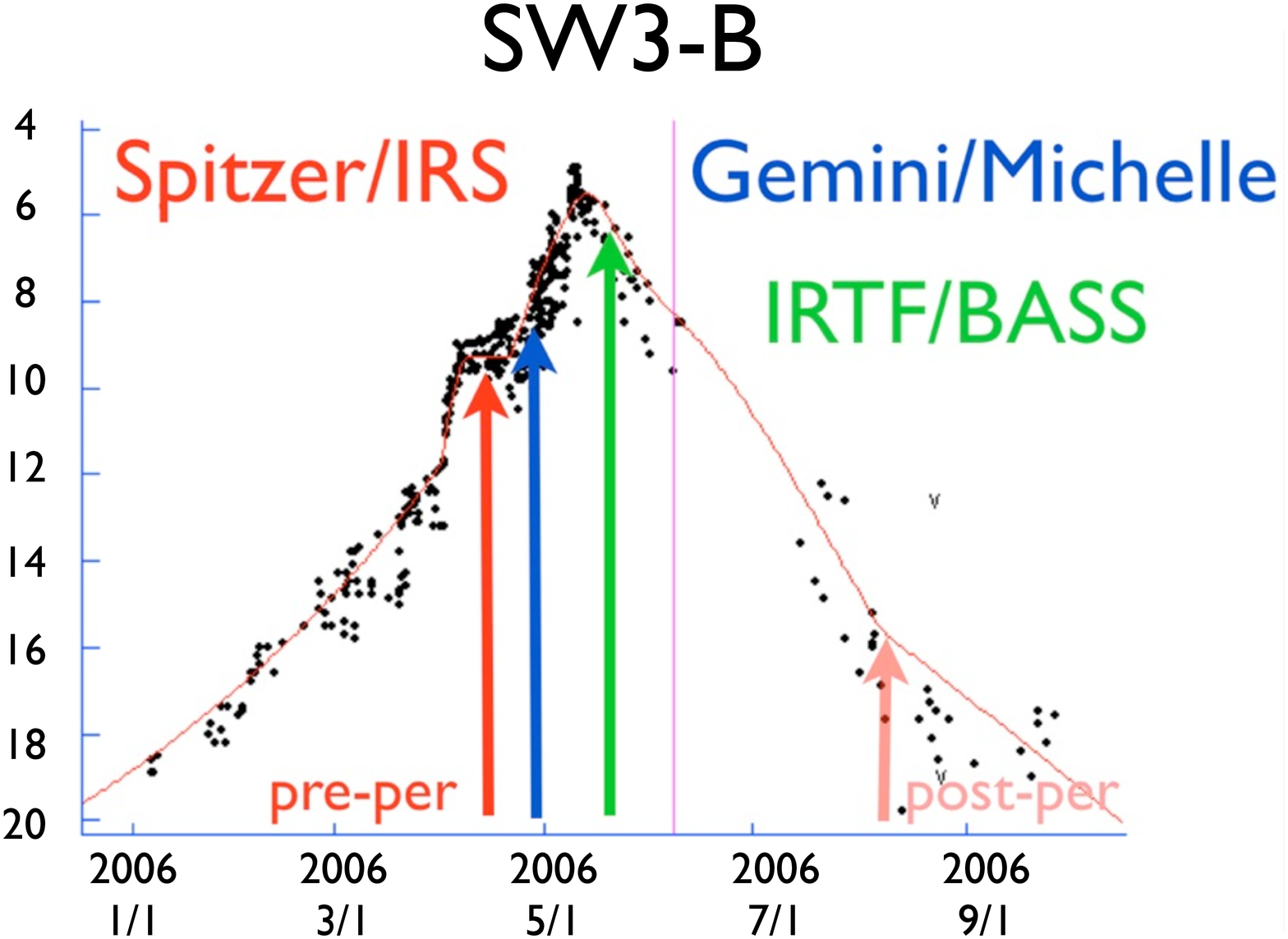} {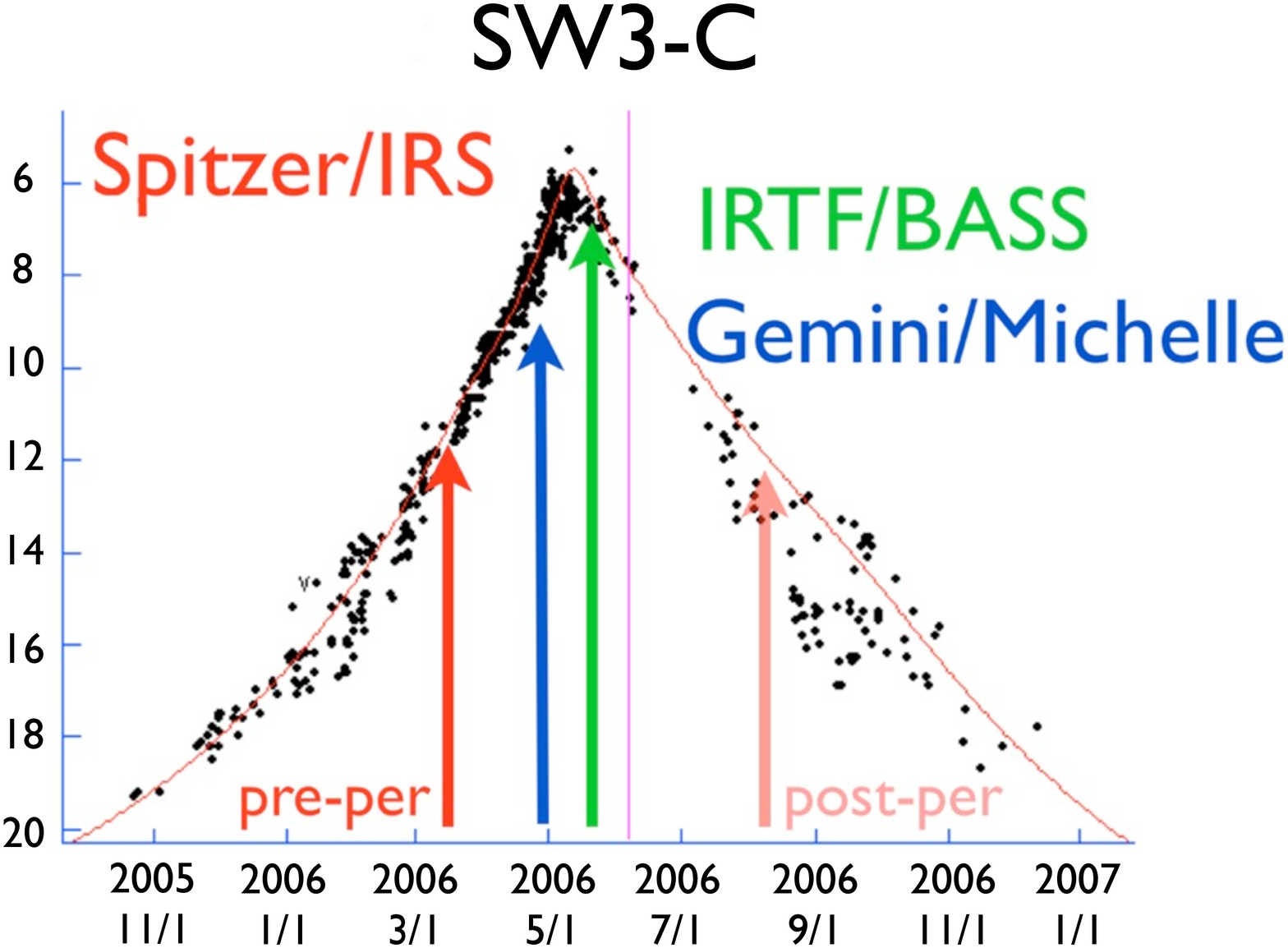}
\caption{The relative timing of the mid-IR observations. The pre-perihelion IRS and the BASS data are discussed in this paper. The Michelle observations are in \citet{har10}. Post-perihelion data were also obtained with the IRS, but are of lower quality than the pre-perihelion data, and are not discussed in this paper.
\label{fig:context}}
\end{figure}

In addition to observations of 73P/Schwassmann-Wachmann 3 obtained with the \textit{Spitzer Space Telescope} we also carried out  ground-based observations of components B and C using the Michelle spectrograph on the Gemini North telescope and on NASA's Infrared Telescope Facility (IRTF) using the SpeX and BASS instruments. The Michelle results are discussed elsewhere \citep{har10}. Here we use the IRTF observations to provide a second estimate of the degree to which scattered solar radiation contaminated the \textit{Spitzer} spectra, determine an estimate of the albedo of the grains, and derive the dust flux parameter \afrho{} commonly used in comet studies. The timing of the \textit{Spitzer}, BASS, and Michelle observations, relative to the light curves of the comet fragments\footnote{Adapted from Sechii Yoshida's comet web site http://www.aerith.net/index.html and used with permission}, is shown in Fig.~\ref{fig:context}.

\subsection{SpeX}

\begin{figure}
\plotone{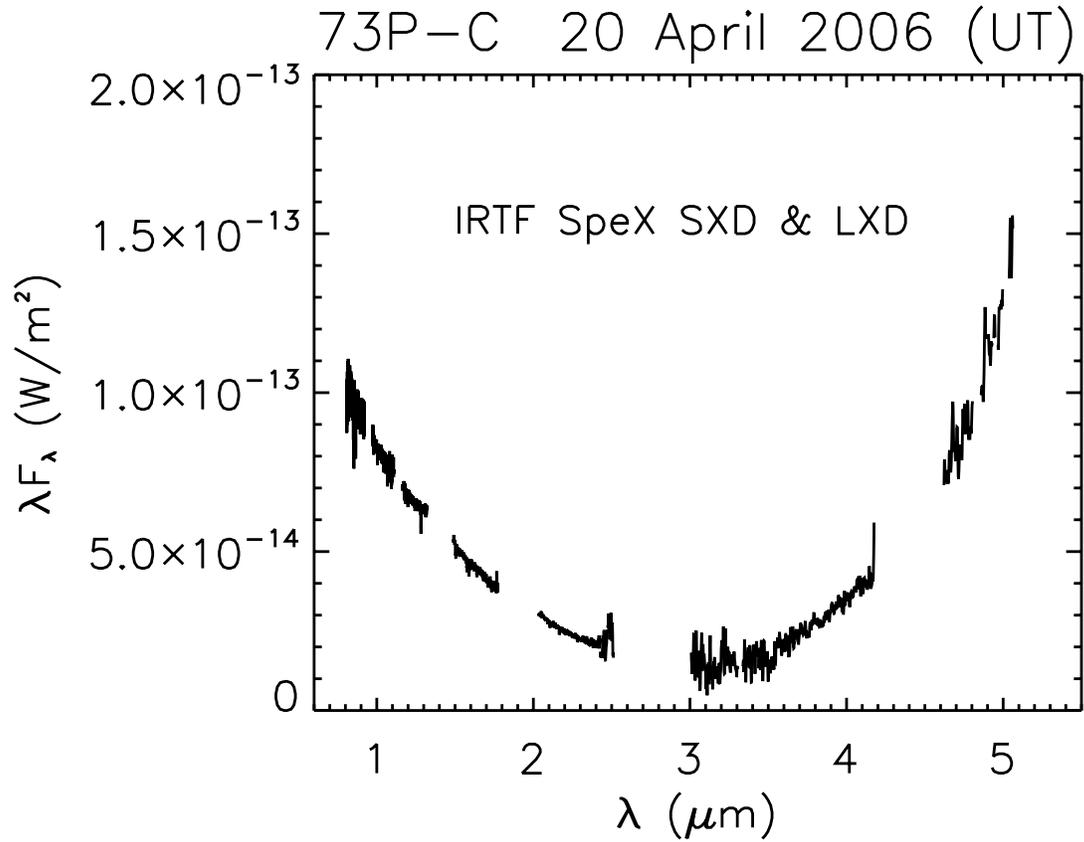} 
\caption{SpeX observations of SW3-C.  The data were obtained using the short wavelength cross-dispersed mode (0.8-2.4~\micron{}; SXD) and the long wavelength cross-dispersed mode (2.3-5.4~\micron{}; LXD2.3) \citep{rayner03}. Going from the shortest to the longest wavelengths, the spectrum transitions from being dominated by scattered solar radiation to thermal emission from the dust grains.
\label{fig:spex}}
\end{figure}

SW3-C was observed on 20 April 2006 (UT) using the SpeX spectrograph on NASA's Infrared Telescope Facility. Observations were made using the cross-dispersed echelle gratings in both short-wavelength mode (SXD) covering 0.8-2.4 \micron{} and long-wavelength mode (LXD) covering 2.3-5.4 \micron{} \citep{rayner03}. All observations were obtained using a 0.8 arcsec wide slit, corrected for telluric extinction and flux calibrated. Because no suitable A0V star was observed close enough in airmass and hour angle to the comet observations, we used data obtained on HD 25152 on a later date with similar sky conditions, airmass, and hour angle. The data were reduced using the Spextool software \citep{vacca03,cushing04} running under IDL.  Details of the observations are listed in Table 4, and the resultant spectrum is shown in Fig.~\ref{fig:spex}. In this figure, the dropping flux density at the shorter wavelengths, followed by the rise at longer wavelengths, illustrates the transition from scattered solar radiation to thermal re-emission by the cometary dust.

\subsection{BASS}

\begin{figure}
\plotone{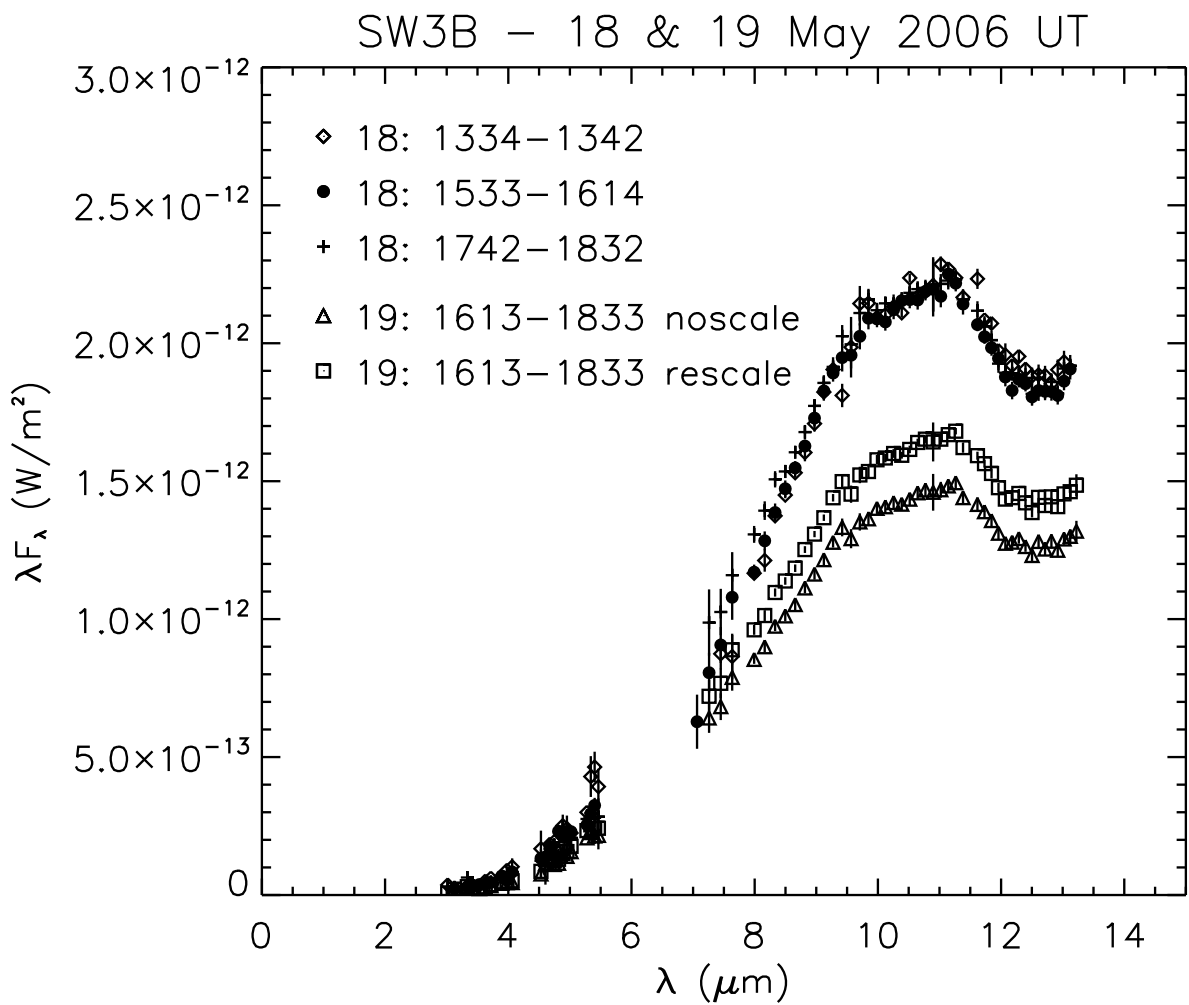} 
\caption{BASS observations of 73P-B. The observations have been broken down into individual sets that were interleaved with similar observations of SW3-C and flux calibration standard stars. The data on 19 May labeled ``noscale'' are an unweighted mean of all of the individual data sets during that time interval. Because errors in guiding, particularly during the daytime, tend to cause the throughput of the light to diminish (such errors never add light to the beam) the true signal may be somewhat larger than a simple average would indicate. Here ``rescale'' is an example of scaling the individual data sets with the highest one (integrating over the same wavelength range for each one). The difference between the two is a rough measure of the systematic uncertainty in the absolute flux density of the comet data, measured though the same fixed entrance aperture. \
\label{fig:b1819}}
\end{figure}

\begin{figure}
\plotone{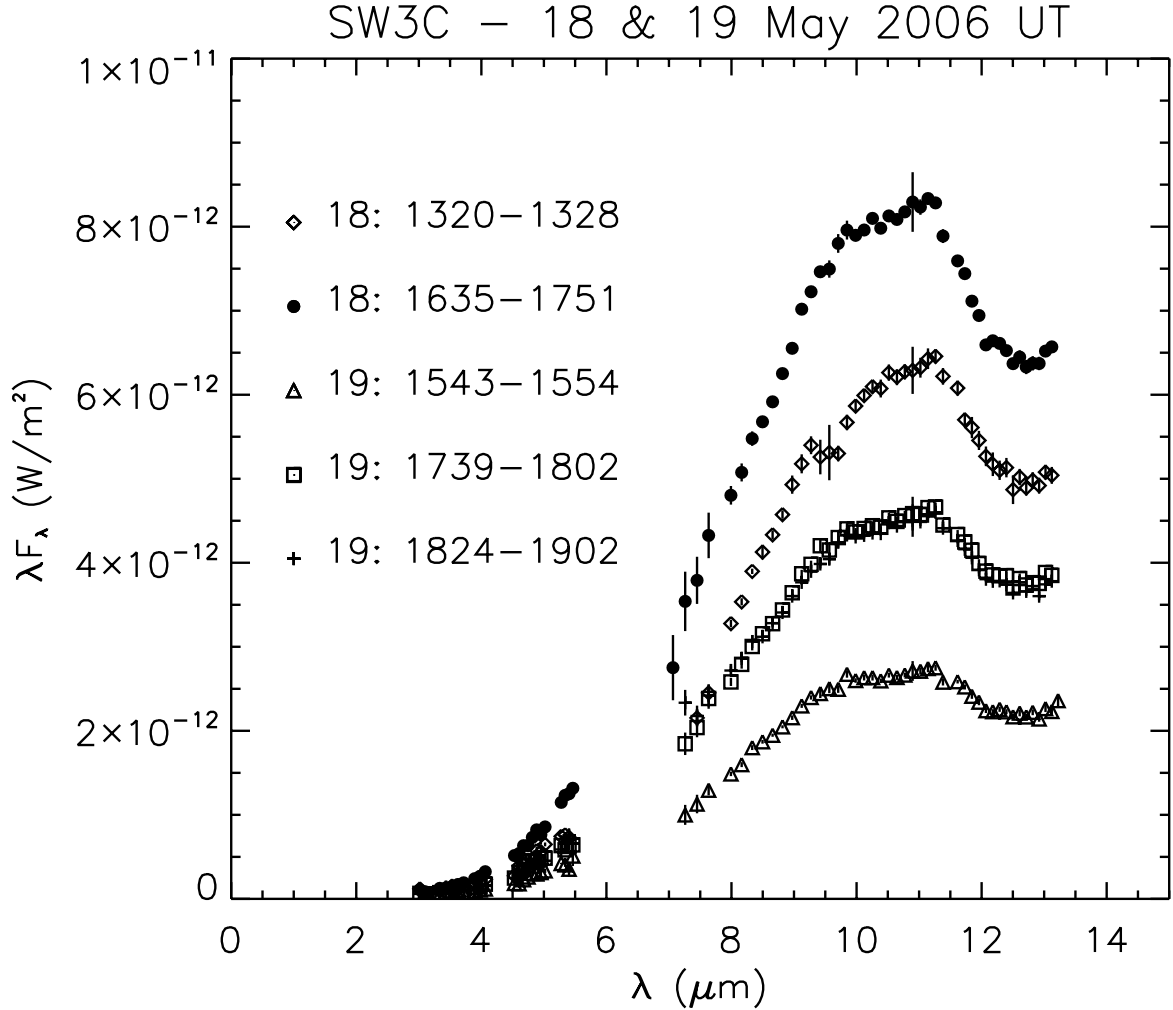}
\caption{BASS observations of SW3-C.  In addition to the changes seen between 18 and 19 May, the data suggest significant intra-night changes. Note that twilight began at $\sim$ 1500 UT and sunrise at $\sim$ 1600 UT, when guiding was done by centering the comet in the aperture to maximize the signal data sets. For component B all three data set form 18 May are identical, despite the second one being obtained in twilight and the third over 2 hours after sunrise. This suggests that the differences seen here in C are likely real, and not due to centering errors during daylight hours.
\label{fig:c1819}}
\end{figure}

Both B and C components were observed with The Aerospace Corporation's Broad-band Array Spectrograph System (BASS) on 18 and 19 May 2006 (UT). BASS uses a cold beamsplitter to separate the light into two separate wavelength regimes. The short-wavelength beam includes light from 2.9-6 $\mu$m, while the long-wavelength beam covers 6-13.5 $\mu$m. Each beam is dispersed onto a 58-element Blocked Impurity Band (BIB) linear array, thus allowing for simultaneous coverage of the spectrum from 2.9-13.5 $\mu$m. The spectral resolution $R = \lambda$/$\Delta\lambda$ is wavelength-dependent, ranging from about 30 to 125 over each of the two wavelength regions \citep{hackwell90}. The details of the observations are listed in Table 5 and the spectra of components B and C are shown in Figs.~\ref{fig:b1819}~\&~\ref{fig:c1819}.

\begin{figure}
\plotone{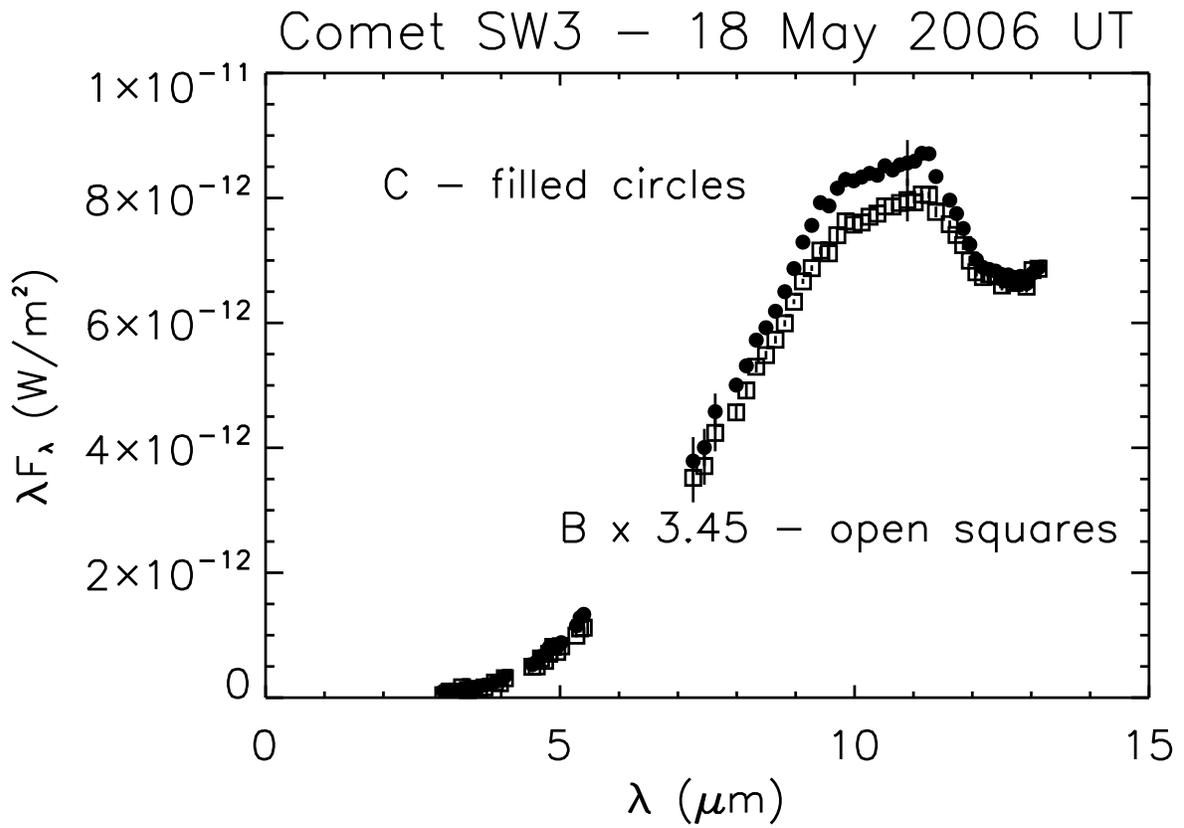}
\caption{Components B and C, normalized to the same flux density at 12.5~\micron{}.  The relative strength of the silicate bands is consistent with the \textit{Spitzer} IRS spectra.
\label{fig:bass060518}}
\end{figure}

The BASS spectra of B and C are nearly indistinguishable, with the exception that the relative strength of the silicate band is slightly stronger in C than in B (Fig.~\ref{fig:bass060518}). This is consistent with what we observed in the \textit{Spitzer} IRS data.

\subsection{BASS CCD} 

On 19 May 2006 (UT) we also obtained observations of SW3-C using the CCD guide camera of BASS and the Blue Continuum (0.44 \micron{}) and Red Continuum (0.71 \micron{}) ``Hale-Bopp'' filters (\citet{farnham00}; hereafter $B_C$ and $R_C$). Observations were calibrated using the flux calibration stars (HD 186408, HD 164852, and HD 170783) for the filters. Observations of 73P-B were unsuccessful due to bright twilight. The details of the CCD observations are listed in Table 6. In addition to the comet filters, we obtained a deeper image using a broad-band Cousins-Kron R filter, but this was not used to extract photometric information. Photometric information was extracted from the $B_C$ and $R_C$ images a circular software aperture matching the entrance aperture of the BASS dewar.

\subsection{Scattered Light Contribution to the mid-IR Spectra}

\begin{figure}
\plotone{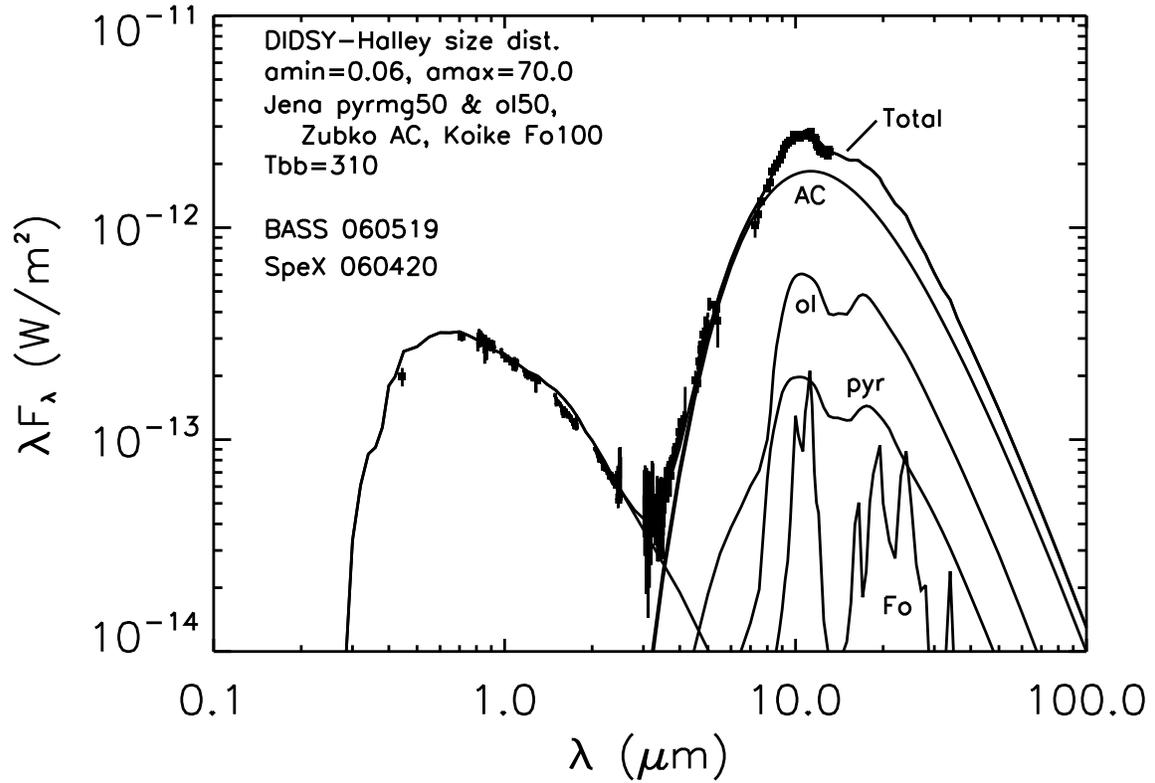}
\caption{Toy model of SW3-C using SpeX and BASS data to estimate the scattering contamination of the IRS data at 5~\micron{}.  The model was used to estimate the degree to which light scattered by the dust contaminate the shortest wavelengths of the IRS data used in the spectral modeling, and to help determine the grain albedos.
 \label{fig:bassmodel}}
\end{figure}

In Fig.~\ref{fig:bassmodel} we show the 3-13 \micron{} spectrum of component C obtained on 10 May 2006 (UT), along with the $B_C$ and $R_C$ flux densities obtained the same night. To this we added the SpeX data, scaled to roughly match both the CCD photometry and mid-IR BASS spectrum. The match in the 3-5 \micron{} region where the SpeX and BASS spectra overlap is only approximate, as they were obtained when the comet was a different heliocentric distances (SpeX - 1.16; BASS - 0.98), and hence slightly different temperatures.

Also shown is a slightly reddened smoothed solar spectrum that approximately fits the combined CCD and SpeX data, plus a simple grain emission model. The model consists of  mixture of four grain components: amorphous olivine and pyroxene\footnote{Strictly speaking, the terms ``olivine'' and ``pyroxene'' refer to pure crystalline end-grroups of the minerals, but we use them here instead of ``amorphous silicates of roughly olivine and pyroxene stoichiometry''}, crystalline olivine, and amorphous carbon. Optical constants were taken from \citet{dor95} for the amorphous silicates with $Mg/Fe =1.0$, that of the crystalline forsterite from \citet{koike03}, and  amorphous carbon (specifically arc discharge from amorphous carbon electrodes in an Ar atmosphere) from \citet{zubko96}. A Mie scattering model of the three amorphous constituents was calculated for solid spheres using size distribution for 1P/Halley determined with the DIDSY instrument on the \textit{Giotto} spacecraft \citep{McDonnell87} with size cutoffs at 0.06 and 70.0 \micron{}. To this was added the crystalline forsterite, and the temperature (for this purpose, assumed to be the same for all the components) adjusted to fit the BASS data. The purpose was not to obtain an independent assessment of the grain composition, but to approximate the scattered light and thermal emission in order to obtain two quantities: the grain albedo, and the degree that scattered light contributes to the wavelengths included in the \textit{Spitzer} spectral modeling. Here we use the fact that B and C are spectroscopically nearly identical, so that the merged data and model for C can be used to estimate the scattered light contamination for both B and C on the dates of the \textit{Spitzer} observations.

For component B  a simple extrapolation of the SpeX data short-wavelength spectral slope to longer wavelengths would suggest 1\% for the scattered light at 5~\micron{}, but looking globally at the SpeX+BASS observations would imply that it is closer to 3\%.  For C the fraction of the light at 5~\micron{} that is due to scattered solar photons is probably about 7\% for the data obtained on March 17 (when the dust was cooler than that of \textit{Spitzer} observations of B). \citet{reach09} put it at 6\% in the 4.5~\micron{} IRAC image, obtained on  March 28. While the solar spectrum used is slightly reddened to approximate the flux levels in the Hale-Bopp filters, the value of 7\% assumes that the grains are becoming a little bit ``blue'' in the infrared, consistent with \citet{jm86}. If the grains are ``gray'' scatterers at all wavelengths, the number is higher, and possibly closer to 10.7\%. The scattered light component will most heavily affect the grain component that dominates at the shortest wavelength, which in our models is the amorphous carbon. However, even a 10\% change in the derived abundances does little to change the fact  that the carbon abundance in the dust of SW3 is higher than the other comets to which the same spectral models has been applied.

\subsection{Grain Albedo}

For this simple model, the bolometric albedo (net integrated scattered light divided by the sum of the scattered light and thermal emission) was $\sim$0.13. At the phase angles that these observations were made (60$^{\circ}$-80$^{\circ}$) this is somewhat lower than many comets, although similar to the curve for C/1980 Y1 (Bradfield) from Fig 1 of \citet{kolokolova04}. Note that this definition of the albedo is the same as that of \citet{gn92}, sometimes called the \textit{bolometric} albedo. This differs from  that of \citet{hn89} and \citet{att86}, which is the \textit{geometric} albedo, usually multiplied by a normalized phase function. The latter are generally much smaller than the bolometric albedo. For a more complete discussion of these terms, see \citet{hn89}.

\subsection{Variability}

We observed both components with BASS over two nights. The observations of B and C were interleaved with each other, with observation of flux calibration stars, and with the CCD imaging. These provided three sets of data on B spanning 5 hours on 18 May, and a single set of observations on 19 May. For C we had two such sets on 18 May spanning 4.5 hours and two sets on 19 May spanning 1.7 hours.

For the 18 May observations of B, the resultant flux levels were identical to within the uncertainties, despite the fact that the second set of observations were obtained after the start of twilight, and the third many hours after sunrise. In both cases, the centering of the comet in the BASS entrance aperture was frequently checked by adjusting the centering to maximize the signal. The simplest explanation for the agreement of the three data sets is that B was stable during these observations. It is unlikely that errors in centering and tracking were precisely cancelled out by source variability, and gives us confidence that the observed changes in brightness in our daytime observations of either comet are likely real. The 30\% drop on brightness on the following night seems large at first sight, but such rapid changes were observed in the H$_{2}$O production rates in B reported by \citet{dr07}. Our BASS observations were obtained closer to the major breakup event of B than the observations of \citet{dr07}, when large variations might be expected.

Contrary to our expectations, C seem to exhibit greater changes in mid-IR emission than B. If a significant fraction of C were ``old'' surface that survived from before the original 1995 breakup, rotational modulation of ``old'' and ``new'' surface to solar heating might explain this behavior. A similar behavior is seen in the data obtained at shorter wavelengths.

Optical observations of comet dust can be expressed in terms of the
product \afrho{}, where $A(\theta)$ is the grain albedo as a function
of phase angle, $f$ is the dust filling factor, and $\rho$ is the
radius of the aperture used to measure the flux \citep{ahearn84},
typically expressed in units of cm.  For comae in an isotropic
steady-state outflow, or a coma that would otherwise follow a
$\rho^{-1}$ surface brightness profile, \afrho{} is independent of
$\rho$.  \afrho{} is used as a proxy for comet dust production rates,
but also depends on the comet's grain size distribution, and dust
scattering properties.

\begin{figure}
 \plotone{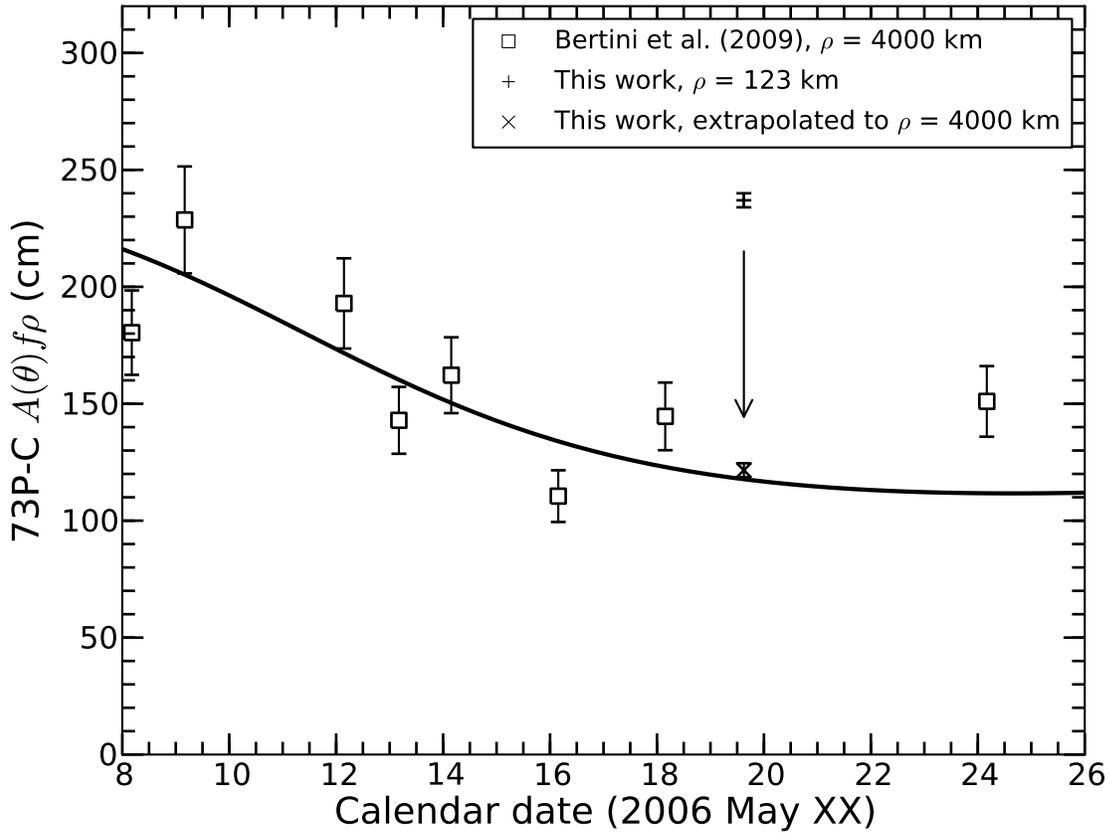}
  \caption{\afrho{} versus observation date for SW3-C.  The
  1\farcs7 radius aperture data is shown as a $+$.  The $R_C$ filter's \afrho{}
   extrapolated to a 4000~km aperture is shown as a $\times$ (see
 for details).  The squares are Johnson $R$-band \afrho{}
   measurements from \citet{bertini09}, along with a qualitative trend line.}
  \label{fig:afrho}
\end{figure}

For SW3-C, the flux density in the $R_C$ filter was F$_{\lambda}$=4.28 x 10$^{-13}$ W m$^{-2}$\micron{}$^{-1}$ ($\lambda$F$_{\lambda}$=3.05 x 10$^{-13}$W m$^{-2}$; F$_{\nu}$=72.5 mJy), and \afrho{}= 237 cm. 
The apertures used were 1\farcs7 (123~km) in radius, equivalent to
the BASS mid-IR entrance aperture.  \citet{bertini09} present Johnson
$R$-band \afrho{} values measured in a 4000~km radius aperture
throughout April and May 2006.  Our $R_C$ filter image does not have
the signal-to-noise ratio required to properly measure \afrho{} in a
4000~km aperture.  To properly compare our $R_C$ value to their
values, we measured the azimuthally averaged surface brightness
profile in our broad-band $R$-band image to determine a scale factor
(0.51) for our $\rho=123$~km measurement.  We compare our corrected value of  \afrho{}=122 cm to those of 
\citet{bertini09} in
Fig.~\ref{fig:afrho}.

To verify fragment C's variability, we examined the \afrho{} data in
Fig.~\ref{fig:afrho}.  A substantial day-to-day variability is
apparent in the \afrho{} data, although it depends on the
interpretation of the errors given in \citet{bertini09}.
\citet{bertini09} state that their $A(\theta)f\rho$ errors are on the
order of $\pm10$\% (we have used 10\% in Fig.~\ref{fig:afrho}), and
that these are primarily flux calibration errors.  If the flux
calibration error is independent of the night the comet was observed,
then the relative scatter in \afrho{} may be real, and consistent changes seen in the BASS data.

Such modulation would not normally be apparent in a
comet coma because short timescale variability would be averaged out
by large physical apertures.  However, comet SW3 passed very close to
the Earth, and was 0.09$\sim$AU away at the time of the BASS observations.
Photometry in a 1\farcs7 radius aperture is sensitive to 
$123v^{-1}$~s timescales, where $v$ is the outflow speed in km~s$^{-1}$ of
the dust being measured.  Micron and smaller sized dust grains are
ejected at speeds of order 0.1~km~s$^{-1}$ at 1~AU from the Sun
\citep{lisse98}, which corresponds to 20 minute timescales.  Thus,
dust comae variability on rotational period timescales could be easily
observed, especially given the small grain dominated grain size
distributions of comet 73P (dn/da $\propto a^{-4.0}$ and $a^{-3.5}$,
\S3.1.3).

\section{Comparison to Halley \textit{in situ} Results}

\subsection{Sampling of 1P/Halley}

In order to provide a comparison to comet abundance data obtained through direct sampling rather than spectral modeling we have used the results of the three impact-ionization mass spectrometers flown through the coma of 1P/Halley on the \textit{Giotto} spacecraft  (PIA instrument)  and two \textit{VEGA} spacecraft  (PUMA-1 and PUMA-2 instruments). The main results of abundances derived from these instruments are those of \citet{langevin87}, \citet{jck88}, and \citet{lawler89}. Due to the limited data transfer rates, the bulk of the spectra were transmitted in compressed modes, while one out of every 30 were uncompressed. In both cases, ions with energies up to either 150 eV or 50 eV were measured, and these comprised the \textit{short} and \textit{long} spectra, respectively \citep{jk91}. \citet{langevin87} included 2284 compressed spectra, combined from all three instruments. \citet{lawler89} began with the same spectra but removed those sets plagued by excessive noise, high backgrounds, and spectra with double peaks. They carefully examined the data sets for solar isotopic ratios, which helped to eliminate contaminating effects, a procedure not used by \citet{langevin87}. The result was 433 compressed PUMA-1 spectra (284 short and 149 long), 29 uncompressed PUMA-1 spectra, and 32 uncompressed PIA spectra. However, they only tabulated the results of the 433 compressed data sets. \citet{jck88}, used the 74 uncompressed PUMA-1 spectra. These were also checked for reasonable isotopic ratios. While these last data may be the cleanest, they represent a sample size only one-sixth that of \citet{lawler89}. 

\citet{jk91} took the uncompressed spectra of the ions and made an approximate conversion into total atoms by estimating the total yields of the instruments. However, the conversion was probably only accurate to within a factor of two. \citet{jk91} tabulated the \textit{ion} ratios of Si, S, and Fe to  Mg their PUMA-1 data with that of \citet{lawler89}, and the ratios for the ions are quite similar, even to the relative ratios in the short and long spectra separately.

\subsection{Normalization to Si versus Mg}

Throughout the literature on cosmochemistry, various groups have determined the relative abundances of the elements by normalizing their values to (usually) either Mg or Si. The abundances of \citet{jck88} and \citet{jk91} used Mg, as the isotope ratio discriminant used to select the cleanest data were considered to be better for Mg. \citet{lawler89} used Si  for their own ion data, and converted the yield-corrected atomic data of \citet{jck88} into Si-normalized values. The CI chondrite data of \citet{lodders09a,lodders09b} also uses Si. For this reason we have plotted all of the data sets (Halley results, CI results and comet spectral modeling results) with both normalizations. Overall, the abundances derived by \citet{lawler89} are close to or slightly smaller than the solar photospheric values \citep{asplund09}, and the short and long data are also fairly consistent with each other for the heavier species. The data of \citet{jck88} and \citet{jk91} exhibit anomalously high (above solar) values for Si and S when normalized to Mg. By comparison, normalizing to Si shows no excess over solar (except for Ca where the uncertainties are larger than most species). 

Which is to be preferred? The greater scatter in the data of \citet{jk91} may be due to the fact that normalizing to Mg versus Si will cause a scale change, as their Si/Mg ratio was $\sim$2, and the uncertainties in converting the ion yields into atomic abundances were relatively uncertain. It may also be that the 74 spectra they used were not as representative of the Halley dust as the 433 spectra used by \citet{lawler89}. Certainly the confidence in the mean values of the abundances derived are likely to be more representative of the true values for the larger data set, if they were otherwise of comparable quality. For these reasons we believe that the Halley data by \citet{lawler89} and the normalization to Si are to be preferred. This choice doe not affect our main conclusions regarding the abundances of C and O, however.



\begin{deluxetable}{lccc}
\tablecolumns{4}
\tablewidth{0pc}
\tablecaption{\textit{Spitzer} IRS Observing Modules}
\tablehead{
\colhead{Module} & \colhead{Abbreviation} & \colhead{Wavelength Range ($\mu$m)}  & \colhead{Resolving Power ($R = \lambda$/$\Delta\lambda$)}}
\startdata
Short Low 2$^{nd}$ Order & SL2 & 5.2-8.7 & 60-127 \\
Short-Low 1$^{st}$ Order & SL1 & 7.4-14.5 & 60-120  \\
Long-Low 2$^{nd}$ Order & LL2 & 14.0-21.3 & 57-126 \\
Long-Low 1$^{st}$ Order & LL1 & 19.5-38.0 & 58-112 \\
Short-High & SH & 9.9-19.6 & $\approx$ 600 \\
Long-High & LH & 18.7-37.2 & $\approx$ 600 \\
\enddata
\end{deluxetable}

\begin{deluxetable}{lcccc}
\tablecolumns{5}
\tablewidth{0pc}
\tablecaption{\textit{Spitzer} IRS Stare Exposure Times for Comet 73P/Schwassmann-Wachmann 3}
\tablehead{
\colhead{Slit} & \colhead{Ramp Time (sec)}   & \colhead{Cycles}    & \colhead{Total Time (sec)}  &  \colhead{Comments}}
\startdata
SL2  & 240 & 2 & 480 & both nominal \\
SL1  & 60 & 2 & 120 & both nominal  \\
LL2  & 120 & 2 & 240 & B saturated 16-19 $\mu$m \\
LL1  & 120 & 2 & 240 & both nominal \\
\enddata
\end{deluxetable}

\clearpage

\begin{deluxetable}{lccccccc}
\tablecolumns{8}
\tablewidth{0pc}
\tablecaption{\textit{Spitzer} IRS Observations of Comet 73P/Schwassmann-Wachmann 3}
\tablehead{
\colhead{Object} & \colhead{Mode}   & \colhead {UT Date} & \colhead{UT Start Time}     & \colhead{Exposure (sec)}  &  \colhead{Comments} 
 }
\startdata
SW3-C & SL2 & 17 March 2006 & 00:58   & 480 & nominal \\
SW3-C & SL1 & 17 March 2006 & 01:17   & 120 & nominal  \\
SW3-C & LL2 & 17 March 2006 & 02:04   & 240 & nominal \\
SW3-C & LL1 & 17 March 2006 & 02:11  & 240 & nominal \\
SW3-B & SL2 & 17 April 2006 & 14:08   & 480 & nominal \\
SW3-B & SL1 & 17 April  2006 & 14:29   & 120 & nominal  \\
SW3-B & LL2 & 17 April  2006 & 15:44  & 240 & saturated 16-19 $\mu$m \\
SW3-B & LL1 & 17 April  2006 & 15:24  & 240 & nominal \\

\enddata
\end{deluxetable}

\clearpage

\begin{deluxetable}{lccccc}
\tablecolumns{6}
\tablewidth{0pc}
\tablecaption{NASA/IRTF SpeX Observations of Comet 73P/Schwassmann-Wachmann 3 - C}
\tablehead{
\colhead{Object} & \colhead{Mode}$\tablenotemark{a}$   & \colhead {UT Date} & \colhead{UT Time}  & \colhead{Hour Angle}  & \colhead{Airmass}}
\startdata
SW3-C & SXD & 20 April 2006 & 10:56 $-$ 11:47 & -1:01 $-$ -0:21 & 1.01 $-$ 1.04  \\
SW3-C  & LXD & 20 April 2006 &  09:47 $-$ 10:42 & -2:11 $-$ -1:02 & 1.06 $-$ 1.16  \\
HD 25152 & SXD & 22 August 2006 & 14:25 $-$ 14:34  & -1:55 $-$ -1.46 &  1.16 $-$ 1.14    \\
HD 25152  & LXD & 22 August  2006 & 14:11 $-$ 14:20 & -2.09 $-$ -2.00 & 1.17 $-$ 1.19 \\
\enddata
\tablenotetext{a} {SXD = Short wavelength cross-dispersed echelle mode; LXD = Short wavelength cross-dispersed echelle mode}
\end{deluxetable}

\clearpage

\begin{deluxetable}{lccc}
\tablecolumns{4}
\tablewidth{0pc}
\tablecaption{NASA/IRTF BASS 3-13~\micron{} Observations of Comet 73P/Schwassmann-Wachmann 3}
\tablehead{
\colhead{Object} & \colhead {UT Date} & \colhead{UT Time}  & \colhead{Airmass}}
\startdata

SW3-B & 18 May 2006 & 15:33 $-$ 16:14 &  1.13 $-$ 1.06 \\
$\beta$ Peg & 18 May 2006 & 16:20 $-$ 16:30 & 1.06 $-$ 1.05 \\
SW3-C & 18 May 2006 & 16:35 $-$ 18:30 &  1.05 $-$ 1.03 \\
$\beta$ Peg & 18 May 2006 & 18:37 $-$ 18:45 & 1.04 \\
$\beta$ Peg & 19 May 2006 & 15:30 $-$ 15:33 & 1.15 \\
SW3-C & 19 May 2006 & 15:43 $-$ 15:54 &  1.15 $-$ 1.18 \\
SW3-B & 19 May 2006 & 16:13 $-$ 17:32 & 1.01 $-$ 1.08 \\
SW3-C & 19 May 2006 & 17:39 $-$ 18:02 & 1.02  \\
$\beta$ Peg & 19 May 2006 & 18:07 $-$ 18:18 & 1.02 \\
SW3-C & 19 May 2006 & 18:24 $-$ 19:02 & 1.04 $-$ 1.07  \\

\enddata
\end{deluxetable}

\clearpage

\begin{deluxetable}{lccc}
\tablecolumns{4}
\tablewidth{0pc}
\tablecaption{NASA/IRTF BASS CCD Hale-Bopp Blue Continuum (B$_C$; 0.44~\micron{}) \& Red Continuum (R$_C$; 0.71~\micron{}) Filter Observations of Comet 73P/Schwassmann-Wachmann 3}
\tablehead{
\colhead{Object} & \colhead {UT Date} & \colhead{UT Time}  & \colhead{Airmass}}
\startdata
HD 164852 & 19 May 2006 & 14:30 $-$ 14:33 & 1.13 \\
HD 170783 & 19 May 2006 & 14:40 $-$ 14:43 & 1.13 \\
73P-C & 19 May 2006 & 14:55 $-$ 15:10 &  1.34 $-$ 1.30 \\
\enddata
\end{deluxetable}

\clearpage

\begin{deluxetable}{lcccccc}
\tablecolumns{7}
\tablewidth{0pc}
\tabletypesize{\scriptsize}
\tablecaption{Composition of the Best-Fit Model to the \textit{Spitzer} IRS Spectrum of SW3-C}
\tablehead{
\colhead{Species} & \colhead {Weighted\tablenotemark{\ a}} & \colhead{Density}  & \colhead{M.W.}  &  \colhead{N$_{moles}$\tablenotemark{b}} & \colhead{Model T$_{max}$\tablenotemark{c}}  & \colhead{Model $\chi^{2}_{\nu}$\tablenotemark{\ d}}  \\
 \colhead{} &  \colhead{Surface Area} &  \colhead{(g cm$^{-3}$)}  &  \colhead{} &  \colhead{(rel.)} &  \colhead{($^{\circ}$K)}&  \colhead{if not included} }
\startdata
\bf{Olivines}  \\
Amorph Olivine (MgFeSiO$_{4}$) & 0.00 & 3.6 & 172 & 0.00 & 225 & 1.01 \\
Forsterite (Mg$_{2}$SiO$_{4}$) & 0.13 & 3.2 & 140 & 0.30 & 225 & 7.98 \\
Fayalite (Fe$_{2}$SiO$_{4}$) & 0.02 & 4.3 & 204 & 0.04 & 225 & 1.21 \\
\\
\bf{Pyroxenes} \\ 
Amorph Pyroxene (MgFeSi$_{2}$O$_{6}$) & 0.34 & 3.5 & 232 & 0.51 & 225 & $ >$100 \\
FerroSilite (Fe$_{2}$Si$_{2}$O$_{6})$ & 0.00 & 4.0 & 264 & 0.00 & 200 & 1.01 \\
Diopside (CaMgSi$_{2}$O$_{4}$) & 0.02 & 3.3 & 216 & 0.03 & 225 & 1.28 \\
OrthoEnstatite (Mg$_{2}$Si$_{2}$O$_{6}$) & 0.00 & 3.2 & 200 & 0.00 & 225 & 1.01 \\
\\
\bf{Phyllosilicates} \\
Smectite Nontronite & 0.03 & 2.3 & 496 & 0.01 & 225 & 1.74 \\
Na$_{0.33}$Fe$_{2}$(Si,Al)$_{4}$O$_{10}$(OH)$_{2}$*3H$_{2}$O \\
\\
\bf{Carbonates} \\
Magnesite (MgCO$_{3}$) & 0.03 & 3.1 & 84 & 0.11 & 225 & 1.41 \\
Siderite (FeCO$_{3}$) & 0.00 & 3.9 & 116 & 0.00 & 225 & 1.01 \\
\\
\bf{Metal Sulfides} \\
Niningerite (Mg$_{30}$Fe$_{70}$S) & 0.06 & 4.5 & 78 & 0.35 & 225 & 3.43 \\
\\
\bf{Water} \\
Water Ice (H$_{2}$O) & 0.15 & 1.0 & 18 & 0.83 & 200 & 17.6 \\
Water Gas (H$_{2}$O) & 0.02 & 1.0 & 18 & 17 & 200 & 1.04 \\
\\
\bf{'Organics'}
\\
Amorph Carbon (C) & 0.53 & 2.5 & 12 & 11 & 280 & $>$100 \\
PAH (C$_{10}$H$_{14}$) & 0.00 & 1.0 & $<$178$>$ & $<$0.008 & NA & 1.01 \\
\enddata
\tablenotetext{a}{Surface Area Weighting(i); A $\pm$ 25\% 2-$\sigma$ error ellipse for the relative surface area of each species was 
determined by noting the fractional amount of change required to increase the 
reduced chi-squared value to just above the 95\% confidence limit.}
\tablenotetext{b}{N$_{moles} \sim$Density(i)/Molecular Weight(i)}
\tablenotetext{c}{$LTE$ at heliocentric distance of 1.47 AU $=$ 233 K}
\tablenotetext{d}{Total best fit model $\chi^{2}_{\nu} =$ 0.99;  95\% confidence limit = 1.13}
\end{deluxetable}

\clearpage

\begin{deluxetable}{lcccccc}
\tablecolumns{7}
\tablewidth{0pc}
\tabletypesize{\scriptsize}
\tablecaption{Composition of the Best-Fit Model to the \textit{Spitzer} IRS Spectrum of SW3-B}
\tablehead{
\colhead{Species} & \colhead {Weighted\tablenotemark{\ a}} & \colhead{Density}  & \colhead{M.W.}  &  \colhead{N$_{moles}$\tablenotemark{b}} & \colhead{Model T$_{max}$\tablenotemark{c}}  & \colhead{Model $\chi^{2}_{\nu}$\tablenotemark{\ d}}  \\
 \colhead{} &  \colhead{Surface Area} &  \colhead{(g cm$^{-3}$)}  &  \colhead{} &  \colhead{(rel.)} &  \colhead{($^{\circ}$K)}&  \colhead{if not included} }
\startdata
\bf{Olivines}  \\
Amorph Olivine (MgFeSiO$_{4}$) & 0.04 & 3.6 & 172 & 0.08 & 245 & 3.67 \\
Forsterite (Mg$_{2}$SiO$_{4}$) & 0.09 & 3.2 & 140 & 0.21 & 245 & 6.86 \\
Fayalite (Fe$_{2}$SiO$_{4}$) & 0.00 & 4.3 & 204 & 0.00 & 245 & 0.99 \\
\\
\bf{Pyroxenes} \\ 
Amorph Pyroxene (MgFeSi$_{2}$O$_{6}$) & 0.28 & 3.5 & 232 & 0.42 & 245 & $ >$100 \\
FerroSilite (Fe$_{2}$Si$_{2}$O$_{6})$ & 0.00 & 4.0 & 264 & 0.00 & 200 & 0.99 \\
Diopside (CaMgSi$_{2}$O$_{4}$) & 0.04 & 3.3 & 216 & 0.06 & 245 & 2.50 \\
OrthoEnstatite (Mg$_{2}$Si$_{2}$O$_{6}$) & 0.00 & 3.2 & 200 & 0.00 & 245 & 0.99 \\
\\
\bf{Phyllosilicates} \\
Smectite Nontronite & 0.04 & 3.2 & 496 & 0.00 & 245 & 0.99 \\
Na$_{0.33}$Fe$_{2}$(Si,Al)$_{4}$O$_{10}$(OH)$_{2}$*3H$_{2}$O \\
\\
\bf{Carbonates} \\
Magnesite (MgCO$_{3}$) & 0.04 & 3.1 & 84 & 0.15 & 245 & 1.85 \\
Siderite (FeCO$_{3}$) & 0.03 & 3.9 & 116 & 0.10 & 245 & 1.65 \\
\\
\bf{Metal Sulfides} \\
Niningerite (Mg$_{30}$Fe$_{70}$S) & 0.01 & 4.5 & 78 & 0.06 & 245 & 1.04 \\
\\
\bf{Water} \\
Water Ice (H$_{2}$O) & 0.06 & 1.0 & 18 & 0.33 & 200 & 3.50 \\
Water Gas (H$_{2}$O) & 0.14 & 1.0 & 18 & 120 & 200 & 1.93 \\
\\
\bf{'Organics'}
\\
Amorph Carbon (C) & 0.65 & 2.5 & 12 & 14 & 300 & $>$100 \\
PAH (C$_{10}$H$_{14}$) & 0.04 & 1.0 & $<$178$>$ & $<$0.023 & NA & 1.29 \\
\enddata
\tablenotetext{a}{Surface Area Weighting(i); A $\pm$ 25\% 2-$\sigma$ error ellipse for the relative surface area of each species was 
determined by noting the fractional amount of change required to increase the 
reduced chi-squared value to just above the 95\% confidence limit.}
\tablenotetext{b}{N$_{moles} \sim$Density(i)/Molecular Weight(i)}
\tablenotetext{c}{$LTE$ at heliocentric distance of 1.20 AU $=$ 257 K}
\tablenotetext{d}{Total best fit model $\chi^{2}_{\nu} =$ 0.99;  95\% confidence limit = 1.13}
\end{deluxetable}

\clearpage

\clearpage

\clearpage

\clearpage
 
\clearpage

\clearpage

\clearpage

\clearpage

\clearpage

\clearpage

\clearpage

\clearpage

\clearpage

\clearpage

\clearpage

\clearpage

\clearpage

\clearpage

\clearpage

\clearpage

\end{document}